\documentclass[11pt]{article}
\usepackage{graphics}
\usepackage{psfrag}
\usepackage{epsfig}
\usepackage{psfig}
\usepackage{epsf}
\usepackage{float}
\usepackage{amssymb,stmaryrd,latexsym}
\newcommand{\beq}{\begin{equation}}

\newcommand{\eeq}{\end{equation}}
\newcommand{\beqa}{\begin{eqnarray}}
\newcommand{\eeqa}{\end{eqnarray}}

\begin{document}

\hfill {\small FZJ--IKP(TH)--2005--08, HISKP-TH-05/06}

\vspace{1in}

\begin{center}

{\Large\bf Precision calculation of \boldmath{$\gamma d\to \pi^+ nn$} \\
\vspace{0.05cm}
within chiral perturbation theory}

\end{center}

\vspace{.3in}

\begin{center}

{\large V. Lensky$^{1,2}$, V. Baru$^{2}$, J.~Haidenbauer$^1$, C.~Hanhart$^1$,

A.E.~Kudryavtsev$^2$, and U.-G. Mei\ss ner$^{1,3}$
}

\bigskip

 $^1$Institut f\"{u}r Kernphysik, Forschungszentrum J\"{u}lich GmbH,\\ 
 D--52425 J\"{u}lich, Germany \\

\bigskip 

 $^2$Institute of Theoretical and Experimental Physics,\\
 117259, B. Cheremushkinskaya 25, Moscow, Russia\\

\bigskip 

 $^3$Helmholtz-Institut f\"{u}r Strahlen- und Kernphysik (Theorie), 
Universit\"at Bonn\\ 
 Nu{\ss}allee 14-16, D--53115 Bonn, Germany 
\end{center}

\vspace{.6in}

\thispagestyle{empty} 

\begin{abstract}
\noindent 
The reaction $\gamma d\to \pi^+ nn$ is calculated up to
order $\chi^{5/2}$ in chiral perturbation theory, where $\chi$ denotes the
ratio of the pion to the nucleon mass.  Special emphasis is put on the
role of nucleon--recoil corrections that are the source of 
contributions with fractional power in  $\chi$. Using the known near threshold 
production amplitude for $\gamma p\to \pi^+ n$  as the only input,
the total cross section for $\gamma d\to \pi^+ nn$ is described very well.
A conservative estimate suggests that the theoretical uncertainty 
for the transition operator amounts to 3 \% for the computed amplitude near threshold.
\end{abstract}

\vfill

\pagebreak


\section{Introduction}

Low--energy meson--nucleus reactions are of great  interest for
they are one of the best tools to deepen our understanding of the nuclear
few--body problem. In particular, reaction involving 
 the lightest member of the Goldstone octet, i.e. the pion, are subject of
special experimental and theoretical efforts since they can be treated within 
chiral perturbation theory (ChPT). 
In this scheme high accuracy calculations in connection
with reliable error estimates are possible. 
Indeed, since the pioneering works by Weinberg \cite{sweinpi} and Gasser
and Leutwyler \cite{gasleu}
ChPT has developed into a powerful tool for
investigations of the $\pi\pi$ \cite{pipi}, $\pi N$ \cite{fettes} as well as few
nucleon systems \cite{fewN}. In addition, it was also Weinberg who pointed out
how to calculate in an equally controlled way pion scattering off as well as
inelastic reactions on nuclei \cite{swein1}. 
There is also a large amount of work in the literature on neutral pion 
production (see Refs. \cite{gammapi0,eepi0} and references therein) as well as on Compton
scattering
(see Refs. \cite{compton} and references therein)
on the deuteron within ChPT.

In this paper we will present, for the first time, a complete calculation
within  ChPT up to order $\chi^{5/2}$ for the
reaction $\gamma d\to\pi^+nn$, where the expansion parameter for the chiral
expansion is denoted by $\chi=m_\pi/M_N$. Here $m_\pi$ ($M_N$) denotes the pion
(nucleon) mass.  Data on this channel exist for excess energies $Q=\sqrt{s}-(2M_N+m_\pi)\le 20$ MeV 
\cite{MIT} which allow us to verify our results. The calculation is of high
theoretical interest, because it provides an important test for our
understanding of those aspects of $\pi NN$ dynamics that are relevant for pion
production reactions on the deuteron. That understanding is a prerequisite for
the reliable extraction of the pion photoproduction amplitude on the neutron,
commonly done from corresponding deuteron data, but also for the determination
of the $nn$ scattering length from $\pi^+$ production data.

In view of the high accuracy of the data and also because of the high
reliability required for the extraction of the above mentioned quantities
involving  neutrons it is now time to critically investigate (and
avoid whenever possible) the approximations traditionally used in pion
reactions on few--nucleon systems. Here we will focus on approximations to the
pion rescattering contribution as they are commonly used in both effective
field theory calculations (see Ref.~\cite{daniel} and references therein) as
well as phenomenological calculations (see Ref.~\cite{levchuk} and references
therein).  Especially for the effective field theory calculations one might
wonder how recoil corrections could be an issue, for the formalism allows for
a rigorous expansion in $m_\pi/M_N$ that should build up the recoil
corrections perturbatively.  However, it was stressed recently \cite{rec} that
the $\pi NN$ threshold introduces non--analyticities in the transition
operators that call for special care: instead of being suppressed by one power
in $m_\pi/M_N$ compared to the formally leading rescattering contributions
(static term), as one might expect naively, the nucleon recoil terms turn out
to scale as $\sqrt{m_\pi/M_N}$ relative to the static term as will be demonstrated in this paper. 
The only publication we are aware of, where the
recoil corrections were treated properly for the near threshold region, is a
phenomenological calculation for $\gamma d\to \pi^0 d$ presented in Ref.~\cite{BenTom} 
(so far most phenomenological studies concentrated on the
$\Delta$--region, cf. Ref.~\cite{aren} and references therein).

Recently the effect of the nucleon recoil on rescattering processes of pions
in $\pi d$ scattering was studied \cite{rec,rekalo,bkt} (for previous
investigations on the role of the nucleon recoil see Refs. \cite{historic}).  In particular, in
Ref.~\cite{rec} we demonstrated that, at least for the $\pi d$ system, the
nucleon recoil can be neglected as long as the two--nucleon intermediate state
is Pauli forbidden, while the pion is in flight. Thus, in this case the static
approximation for the pion exchange is justified. However, as soon as the
two--nucleon state is Pauli allowed, the nucleon recoil has to be included. In
this case it turned out that the whole rescattering contribution (i.e. static
term + recoil corrections) practically canceled completely. It should be
stressed that for $\pi d$ elastic scattering the Pauli allowed two--nucleon intermediate 
states are anyhow suppressed by chiral symmetry.

\begin{figure}[t]
\begin{center}
\epsfig{file=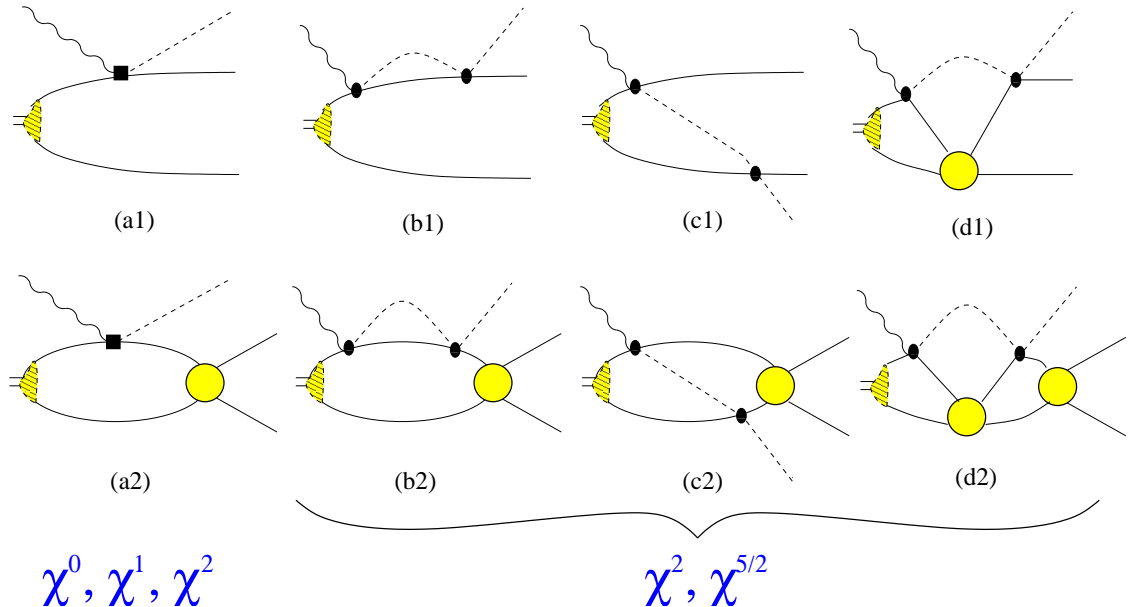, height=8cm, angle=0}
\caption{Diagrams for $\gamma d\to\pi^+nn$. Shown
  are one--body terms $\bigl ($diagram (a) and (b) $\bigr)$, as well as the
  corresponding rescattering contribution (c)---all without and with final
  state interaction. The need to include the $NN$ interaction
  non--perturbatively implies also the inclusion of diagrams with the two
  nucleon pair interacting, while a pion is in flight---this class is shown in
  diagrams (d). Solid straight, wiggly and dashed lines denote
  nucleons, photons and pions, in order.  Filled squares and ellipses  stand for the
  various vertices, the hatched area shows the deuteron wave function and the
  filled circle denotes the $nn$ scattering amplitude.  Crossed terms (where the external lines
  are interchanged) are not shown explicitly. The tree level $\gamma p\to
  \pi^+n$ vertex, as it appears in diagram (a1) and (a2), contributes at leading
  order (order $\chi^0$), and order $\chi^1$ and $\chi^2$, depending on the
  one--body operator used.  Loops start to contribute at order $\chi^2$, the
  corresponding recoil corrections enter at order $\chi^{5/2}$.}
\label{diagam}
\end{center}
\end{figure}

In this paper we will study the reaction $\gamma d\to \pi^+ nn$ with special
emphasis on the abovementioned recoil corrections.  In
this system the selection rules are such that the S-wave two--nucleon intermediate
state during the pion rescattering process is allowed by the Pauli principle.
In addition, once the strength of the one--body
operator is fixed to the reaction $\gamma p\to \pi^+n$, no free parameters
occur in the calculation for $\gamma d\to \pi^+ nn$ to the order where the
leading recoil corrections enter and thus we can compare our results to
experimental data directly. At the same time we get a better understanding of
the few--body corrections to $\gamma d\to \pi^0 pn$. This reaction will
eventually allow one to extract the amplitude of $\gamma n\to \pi^0 n$
complementary to using $\gamma d\to \pi^0 d$ discussed in Ref.~\cite{gammapi0}.
Note that up to now no calculation for $\gamma d\to \pi^+ nn$ exists where
the nucleon recoil was properly included.

\begin{table}[t]
\renewcommand{\baselinestretch}{1}
\begin{center}
\vskip 0.2cm
{\large
\begin{tabular}{|c|l|l|l|}
\hline
order & $s$--wave & $p$--wave & $d$--wave\\
\hline
& & & \\
$1$ & $\chi_m^0$ & \phantom{$\chi_m^2\chi_Q, \, \chi_m^{5/2}$} &
 \phantom{$\chi_m^2\chi_Q, \, \chi_m^{5/2}$} \\
& & & \\
$\chi^1$ & $\chi_m^1 , \, \chi_Q^2$ & $\chi_Q$ & \\
& & & \\
$\chi^2$ & $\chi_m^2 , \, \chi_Q^2\chi_m , \, \chi_Q^4$ & $\chi_Q\chi_m$ & $\chi_Q^2$ \\
& & & \\
$\chi^{5/2}$ & $\chi_m^2\chi_Q, \, \chi_m^{5/2}, \, \chi_m^{1/2}\chi_Q^4$ & & \\
& & & \\
\hline
\end{tabular}}
\caption{Pattern of appearance of the expansion parameters
  $\chi_m=m_\pi/M_N$ and $\chi_Q=k_\pi/m_\pi$ on the amplitude level for a
  given order in $\chi$ for the total cross section. The first column shows
  the order parameter $\chi$, whereas the other three columns show the order
  assignments for the amplitudes of the various pion partial waves of
  relevance. Note, as we here consider the total cross section only, different
  partial waves do not interfere---this was used in the order assignment.}
\label{Table_1}
\end{center}
\end{table}

Before going into the details some comments are necessary regarding the
relevant scales of the problem. In the near threshold regime of interest here
(excess energies of at most 20 MeV above pion production threshold) the
outgoing pion momenta are  small compared even  to the pion mass. Thus, in addition
to the conventional expansion parameters of ChPT $m_\pi/\Lambda_\chi$ and
$q_\gamma/\Lambda_\chi$, where $\Lambda_\chi$ denotes the chiral symmetry
breaking scale of order of (and often identified with) the nucleon mass, and
$q_\gamma$ denotes the photon momentum in the center--of--mass system which is of
order of the pion mass, we can also regard $k_\pi/m_\pi$ as small, where
$k_\pi$ denotes the outgoing pion momentum.  In what follows we will
perform an expansion in two parameters, namely
$$
\chi_m = m_\pi/M_N \ \mbox{and} \ \chi_Q = k_\pi/m_\pi \ .$$
Obviously, the
value of the second parameter depends on the excess energy.  The energy regime
of interest to us are excess energies up to 20 MeV. The maximum value of
$\chi_Q$, $\chi_Q^{max}=\sqrt{2Q/m_\pi}$, possible at maximum energy is thus
about 1/2. Since this is numerically close to $\sqrt{\chi_m}$ we use the
following assignment for the expansion parameter \beq \chi \sim \chi_m \sim
\chi_Q^2 \ .
\label{expansion}
\eeq In this paper we will concentrate on total cross sections only and
thus---up to one important exception---$\chi_Q$ appears with even powers only.
In table \ref{Table_1} we show the powers of $\chi_m$
and $\chi_Q$ that appear in the amplitude as well as the corresponding order
$\chi$ for the total cross section. The pertinent diagrams will be discussed
in detail below. 
Note that the diagrams with $\pi N$ rescattering (see diagrams (b),
(c) and (d) in Fig.~\ref{diagam}) contribute at order $\chi_m^2$ as well as at 
$\chi_m^2\chi_Q$, $\chi_m^{5/2}$ and at $\chi_m^{1/2}\chi_Q^4$. The origin of the non--integer power of
$\chi$ are the two--body ($\pi N$) and three--body $\pi NN$ singularities.
This issue is discussed in detail in section \ref{cuts}.

The small pion momentum in the exit channel leads to a suppression of
higher pion partial waves. As can be seen from table \ref{Table_1} we need to consider at
most pion $d$--waves (here and in what follows we denote pion partial waves by
small letters and $NN$ partial waves by capital letters).  On the other hand,
at excess energies of 20 MeV the maximum two nucleon momentum in units of the
pion mass is of order 1 and thus there is a priori no suppression of
higher $NN$ partial waves. However, since the $nn$ phase shifts are only sizable
for $S$- and $P$-waves
 at the small energies of relevance, we only
include the $nn$ final state interaction of those partial waves.

To summarize the scope of this work, it aims to improve the existing
calculations for the reaction $\gamma d\to \pi^+ nn$ in the following
important aspects:
\begin{itemize}
\item A first complete ChPT calculation for the reaction $\gamma d\to \pi^+ nn$ up to order
  $\chi^{5/2}$ is presented. At this order loops  contribute and the
  non--perturbative character of the $NN$ interaction calls for the use of
  interacting two nucleon Green's functions (leading to up to 3 loop diagrams
  as shown in diagram (d2) of Fig.~\ref{diagam}).  
\item  The leading nucleon recoil is included without approximation. It 
  enters at order $\chi^{5/2}$.
\item As always in effective field theory studies an estimation of 
the accuracy of the calculation can be given. A conservative estimate points
  at an accuracy of 2 \% for the few--body corrections to the amplitude
near threshold which is of the same order as the uncertainty of 
the input quantity $E_{0+}$---the invariant electric dipole amplitude for
$\gamma p\to \pi^+ n$ at threshold. Adding the two uncertainties in quadrature
we arrive at a total uncertainty of 3 \% for the full transition operator.
In this work we use phenomenological $NN$ wavefunctions and thus we are not in
the position to estimate the uncertainty of the complete matrix elements (see
the corresponding discussion in the summary).
\item For the first time pion $p$-- and $d$--waves as well as the final state
  interaction in the $NN$ $P$--waves are included in a calculation for the
  near threshold regime. Note that the significance of $NN$ $P$--waves even at photon
  energies below 20 MeV was established long ago \cite{Lagrev}, however, so
  far they were only considered as plane waves in the impulse term (thus only
  through diagram (a1) of Fig.~\ref{diagam}). 
\end{itemize}

Our paper is structured as follows: in the next section we briefly review
the central findings of Ref.~\cite{rec}. 
The three--body dynamics is discussed in detail in section 3.
The same formalism as described for $\pi d$ scattering in section 2 is
applied to the reaction $\gamma d\to \pi^+ nn$ in section 4. In section~5 we 
present the results followed by a brief summary. The details of the
calculation for the various
diagrams are given in   appendices.

\section{Remarks on the $\pi d$ system}

\begin{figure}[t]
\begin{center}
\epsfig{file=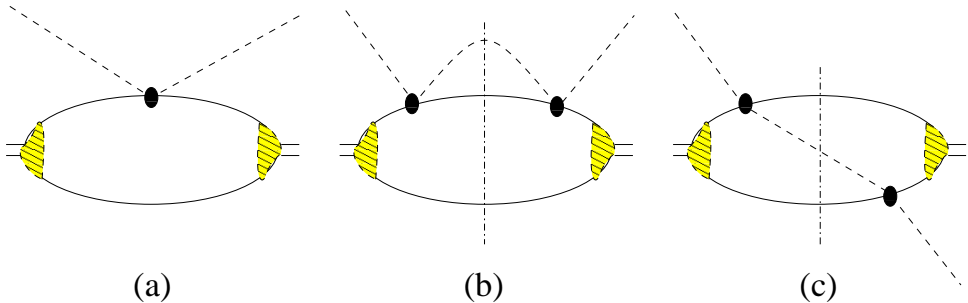, height=4cm, angle=0}
\caption{Typical diagrams for $\pi d$ scattering. Shown are one--body terms
  $\bigl ($diagram (a) and (b), as well as the corresponding
 rescattering   contribution (c)$\bigr)$. 
Crossed terms (where the external pion lines are
 interchanged) are not shown explicitly.}
\label{dia}
\end{center}
\end{figure}

To keep the paper self--contained, in this section we briefly review the
findings of Ref.~\cite{rec}. The relevant diagrams for $\pi d$ scattering 
are shown in Fig.
\ref{dia}. Diagram~(a) denotes the tree level one--body contribution, diagram
(b) the loop correction to the one--body piece and diagram~(c) the rescattering
term.  It should be clear that the Pauli principle calls for a consistent
simultaneous treatment of (b) and (c), for an interchange of the two nucleons
while the pion is in flight (this intermediate state is marked by the
perpendicular line in Fig.~\ref{dia}) transforms one diagram into the other.

The calculations of Ref.~\cite{rec} were based on the effective  $\pi N$ vertex,
\begin{equation}
V^{ba}_{\pi N}=\delta^{ba}g_++\frac i{\sqrt{2}}\epsilon^{abc}\tau^cg_- \ ,
\label{pinpot}
\end{equation}
where $a,b,$ and $c$ are the Cartesian pion indices.
We note that this  vertex 
leads, up to corrections of $O(m_\pi/M_N)$, to
identical results for the $\pi d$ scattering length
as the leading order chirally symmetric $\pi N$ interaction
(the Weinberg--Tomozawa term) if we choose $g_-=-\sqrt{2}m_\pi/(2f_\pi^2)$ and
$g_+=0$. However, to keep the expressions simple we will use the notation of Ref.~\cite{rec}.

The main issue of Ref.~\cite{rec} was to properly isolate the single nucleon
contribution (the one that would be measured in $\pi N$ scattering) from the
few body corrections. It is clear that part of diagram~(b) contributes to the
former and part to the latter.  As outlined in Ref.~\cite{rec} the proper
prescription to separate these two pieces is to add to the tree level
scattering (diagram~(a)) the single nucleon one--loop contribution on the free
nucleon at rest---this sum is the expression for the $\pi N$ scattering
length, $a^{(1-body)}$. The same loop needs to be subtracted from the full
contribution depicted in diagram~(b). This procedure at the same time renders
the expression for the loop finite. This difference is a true two--nucleon
operator. To obtain more symmetric, easier to interpret results
we subtracted from the rescattering piece the expression for the static
exchange, i.e. the contribution from diagram~(c) of Fig.~\ref{dia} calculated
in the static limit.  In order to leave the final result unchanged, this
contribution needs to be added as $a^{(st)}_{LO}$ (and $a^{(st)}_{NLO}$) to the expression for the $\pi
d$ scattering length. Thus, the full result for the $\pi d$ scattering length
reads
\begin{equation}
a=a^{(1-body)}+a^{(st)}_{LO}+a^{(rec)}+a^{(st)}_{NLO} \ ,
\end{equation}
where the individual contributions for the static (st), the recoil (rec) and the NLO
corrections to the static term are given by
\begin{eqnarray}
a^{(st)}_{LO}{=}(g_{+}^2{-}g_{-}^2)I_0 \ ; \ \
a^{(rec)}{=}{g_{+}^2}I_{+}{+}{g_{-}^2}I_{-}  \ ; \ \
a^{(st)}_{NLO}{\simeq}{-}\frac{m_\pi}{M_N}(g_{+}^2{-}g_{-}^2)I_0 \ .
\label{stat}
\end{eqnarray}
The integrals denoted by $I_0$ and $I_\pm$, given explicitly in Ref.~\cite{rec}, 
are evaluated numerically using the deuteron wave functions from
the Bonn potential \cite{bonn}:
$$
I_+= -0.88~I_0 \ , \ \qquad I_-= -0.19~I_0~.
$$
These numbers clearly reflect the claim made above: for the $\pi d$ system the
isovector $\pi N$ interaction, proportional to $g_-$, leads to a Pauli blocked
intermediate state and numerically we find that the recoil corrections lead
only to a 20\% correction (Integral $I_-$ is numerically small, $I_- \ll I_0$.). When the 
NLO pieces mentioned above are added,
the correction resulting from recoil and NLO terms together is only 4\% and
thus the static pion exchange is a good approximation for the total rescattering contribution. 
 On the other hand, the isoscalar $\pi N$
interaction leads to a Pauli allowed intermediate state and in this case the
recoil corrections cancel 90\% of the static exchange ($I_+$ is large and 
negative, $(I_0+I_+)\ll I_0$). The inclusion of the NLO piece in this case
further reduces the total rescattering contribution down to 3 \% of the static
term. Therefore, in this case estimating the total rescattering contribution by
the static exchange only is a very poor approximation.

\section{The role of the $\pi NN$ cuts}
\label{cuts}

\begin{figure}[t]
\begin{center}
\epsfig{file=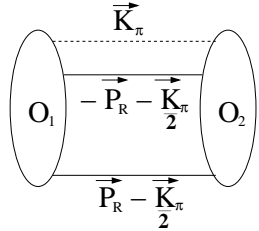, height=3cm, angle=0}
\caption{Notation used in the discussion of the $\pi NN$ intermediate state. The
  ellipses labeled $O_1$ and $O_2$ denote the transition operators for the
  various possible  reactions; e.g. for $\gamma d\to \pi^+ nn$ the operators
  $O_1$ and $O_2$ refer to $\gamma d\to \pi^+ nn$ and $\pi^+ nn\to \pi^+ nn$, respectively.}
\label{cutpic}
\end{center}
\end{figure}

In this  section we discuss the role of the three--nucleon cuts in 
more general terms. 
 All arguments presented for the three--body dynamics apply to pion--nuclear
reactions in general (as, e.g., $\pi d$
scattering), however, in this section we will use only the reaction $\gamma
d\to \pi^+nn$ as illustrative example.

Typical diagrams that contain a $\pi NN$ intermediate state
(cf., e.g., diagram~(c2) of Fig.~\ref{diagam} and the  corresponding Eq.~(\ref{ic2}) in appendix A),
can be cast into the following form by a proper choice of variables (and
dropping terms of higher order in $m_\pi/M_N$)
\beqa I_{\pi NN}(Q)=\int \frac{dK_\pi K_\pi^2 \, dP_R P_R^2}{(2\pi)^6}\frac{f(
  K_\pi^2,P_R^2)}{Q-K_\pi^2/(2m_\pi)-P_R^2/M_N + i0} \ ,
\label{ipinn}
\eeqa
where $K_\pi$ denotes the pion momentum  and $P_R$ denotes
the relative momentum of the two nucleons while the pion is in flight
(c.f. Fig.~\ref{cutpic}).
Using the static approximation means to expand the denominator of the integral
in Eq.~(\ref{ipinn}) in powers of $m_\pi/M_N$ before evaluation of the integral.
 Thus, we get in the static approximation\footnote{Note, we here use the
   phrase 'static approximation' in a quite broad sense in that we also allow
   for the inclusion of correction terms analytic in $\chi$. In
   phenomenological studies those corrections are normally dropped.} 
\beqa I_{\pi NN}^{(static)}(Q)=\int \frac{dK_\pi K_\pi^2 \, dP_R P_R^2}{(2\pi)^6}\frac{f(
  K_\pi^2,P_R^2)}{Q-K_\pi^2/(2m_\pi) + i0} + O\left(\frac{m_\pi}{M_N}\right) \ .
\label{ipinn_st}
\eeqa 
We will demonstrate now analytically (and in section \ref{resadis} numerically) 
that this procedure misses an important contribution to $I_{\pi NN}$.

Above the pion production threshold the denominator
in Eq.~(\ref{ipinn}) has a three--body singularity
that leads to an imaginary part. This imaginary part of $I_{\pi NN}$
 can be calculated by the replacement
$$
(Q-K_\pi^2/(2m_\pi)-P_R^2/M_N+i0)^{-1} \ \to \ -i\pi \delta
(Q-K_\pi^2/(2m_\pi)-P_R^2/M_N) \ .
$$
Thus we find 
\beqa I_{\pi NN}^{(cut)}(Q) = -i\pi m_\pi \int
\frac{dP_R P_R^2}{(2\pi)^6}f(2m_\pi(Q-P_R^2/M_N),
P_R^2)\sqrt{2m_\pi(Q-P_R^2/M_N)} \ .
\label{pinncut}
\eeqa 
From this formula we see that the imaginary parts of the pion loops in
diagram~(b2) and (c2) of Fig.~\ref{diagam} are finite and lead to a strongly
energy--dependent contribution. As long as the  momentum--dependence of the function
$f$ can be neglected, this part of the amplitude grows like $Q^2$, i.e. like
the three--body phase space. 
 These are the contributions at order
$\chi_m^{1/2}\chi_Q^4$ given in table~\ref{Table_1}. Note that the corresponding
amplitudes without final state interaction ((b1) and (c1) of
Fig.~\ref{diagam}) do not have a three--body cut but only a two--body
singularity. Thus their imaginary parts scale as $\chi_m^2\chi_Q$.

On the other hand, the unitarity cut contribution of $I_{\pi NN}^{(static)}$ reads
$$
I_{\pi NN}^{(cut,static)}(Q) = -i\pi m_\pi \sqrt{2m_\pi Q}
\int \frac{dP_R P_R^2}{(2\pi)^6}f(2m_\pi Q,
  P_R^2) \ .
$$
  The most remarkable difference between $I_{\pi NN}^{(cut)}$ and $I_{\pi
    NN}^{(cut,static)}$ is that in contrast to the former the latter scales as
  $\sqrt{Q}$, i.e. like the two body phase space for all diagrams (even for those
  with final state interaction) and therefore shows an energy dependence that
  is completely wrong---and at variance even with perturbative three--body
  unitarity.

For values of $P_R$ with $Q>P_R^2/M_N$ the integral $I_{\pi NN}^{(cut)}(Q)$ 
(cf. Eq.~(\ref{pinncut})) contributes to the
imaginary part of  $I_{\pi NN}(Q)$ (cf. Eq.~(\ref{ipinn})).
To evaluate the contribution of the cut to the real part, the square root needs
to be analytically continued to negative values of its argument through
$$\sqrt{2m_\pi(Q-P_R^2/M_N)} \to i\sqrt{2m_\pi(P_R^2/M_N-Q)} \ .$$
To demonstrate explicitly
the impact of this, let us consider the case $Q=0$. Then we get
\beqa
I_{\pi NN}^{(cut)}(0) = \pi m_\pi \sqrt{\frac{2m_\pi}{M_N}}
\int \frac{dP_R P_R^2}{(2\pi)^6}f(-2m_\pi P_R^2/M_N,
  P_R^2)P_R \ .
\label{cut}
\eeqa
The corresponding expression for $I_{\pi NN}^{(cut, static)}(0)$
vanishes. 
On the other hand, the expression for the leading static approximation
(cf. Eq.~(\ref{ipinn_st}))
 gives at threshold
\beqa
I_{\pi NN}^{(static)}(0) =  -2m_\pi
\int \frac{dP_R P_R^2 dK_\pi}{(2\pi)^6}f(K_\pi^2,
  P_R^2) \ .
\label{static}
\eeqa 
The static approximation Eq.~(\ref{static}) only acquires corrections
analytic in $(m_\pi/M_N)$ and thus  misses
the contribution of Eq.~(\ref{cut}).    
In fact, Eq.~(\ref{cut}) corresponds to the threshold value of the
mentioned non--analytic contribution from the three--body intermediate state
that is dropped in the static approxi{-}mation---the extension to arbitrary
values of $Q$ is straightforward. However, the contribution
from $I_{\pi NN}^{(cut)}$ is significant as one can see
from a naive dimensional analysis where all momenta---even those in the
integral measure---are replaced by their typical values: 
\beqa I_{\pi
  NN}^{(cut)} \sim - \sqrt{\frac{m_\pi}{M_N}} I_{\pi NN}^{(static)} \ ,
\label{scaling} \eeqa
as claimed in the introduction. Therefore, in general the static approximation
is to be avoided!
As one can see, the contributions from the nucleon recoil through
the three--body singularities to both the real and the imaginary 
part of the amplitude appear to be down by
$\sqrt{\chi}$ compared to the leading loop contribution (the static piece). To be more explicit 
they contribute at orders $\chi_m^{5/2}$ and $\chi_m^{1/2}\chi_Q^4$ (cf. table \ref{Table_1}). 
On the other hand, if we expand the propagator in
Eq.~(\ref{ipinn}) before the integration, we only get terms analytic in the
pion mass (order $\chi$ corrections to the static piece) and miss the most prominent correction.
 
Now we are in the position to discuss in more detail the conjecture presented
in Ref.~\cite{rec} as well as in the previous section, namely that in general for all those diagrams, where the
$S$-wave two--nucleon intermediate state that appears while the pion is in flight is
forbidden by the Pauli principle, the various recoil corrections largely
cancel, while in case of Pauli allowed S-wave intermediate states they add
coherently. In the latter case the net effect of the rescattering diagrams was
claimed to be small due to a destructive interference between the recoil
corrections and the leading rescattering contribution (the static term---cf.
Eq.~(\ref{ipinn_st})). We will now discuss the two cases in the light of the
discussion above. The essential observation is that the recoil corrections are
the analytic continuation of the imaginary parts related to on--shell $\pi NN$
intermediate states.

{\it Pauli forbidden intermediate states:} 
in this case the imaginary contributions stemming from the three-body unitarity cut in
diagrams of the type of (b2) and those from the type of (c2) of Fig.~\ref{diagam} 
need to cancel exactly, for the corresponding $\pi NN$ state is
forbidden. As a consequence, there will also be a cancellation for the
analytic continuations (see Eq.~(\ref{pinncut})), i.e. for the
corresponding principal value (PV) integrals, and thus the recoil
corrections necessarily cancel to a large degree.  Since this statement
is based solely on the Pauli principle it must hold for all reactions.

{\it Pauli allowed intermediate states:} contrary to the first option in this
case the corresponding unitarity cut parts of the diagrams  (b2) and (c2) of Fig.
\ref{diagam} add coherently and, as above, the same is to be true for their
analytic continuation. It is not possible to claim in general that the
recoil contribution from the PV integrals completely cancels the whole
static term.
However, as illustrated by the estimate of Eq.~(\ref{scaling}), the recoil corrections tend 
to be of the order of magnitude
of the static term and tend to interfere destructively with it.
Here we refer to the discussion in section \ref{resadis} and to Fig.~\ref{stvsfull}.

As we discussed in the previous section, the leading operator for $\pi N$
scattering acting on a deuteron leads to a $\pi NN$ state, where the $S$--wave
is forbidden for the $NN$--pair  by the Pauli principle. On the other hand, as
will be outlined below, the leading operator for $\gamma N\to \pi N$, when
acting on a deuteron field, leads to a $\pi NN$ state, where the $NN$--pair is
allowed to be in an $S$--wave. 
Thus, in the reactions $\pi d\to \pi d$, $\pi d\to \gamma NN$, and $\gamma
d\to \pi^0d$ the static approximation indeed accounts for a significant
fraction of the few--body corrections, whereas in $\gamma d\to \pi^+ nn$ and
$\gamma d\to \gamma d$, when evaluated near the pion threshold, the static
approximation is expected to work quite poorly.
References for the corresponding calculations are given in the introduction.

\section{The reaction \boldmath{$\gamma d\to \pi^+ nn$}}

\begin{figure}[t]
\begin{center}
\epsfig{file=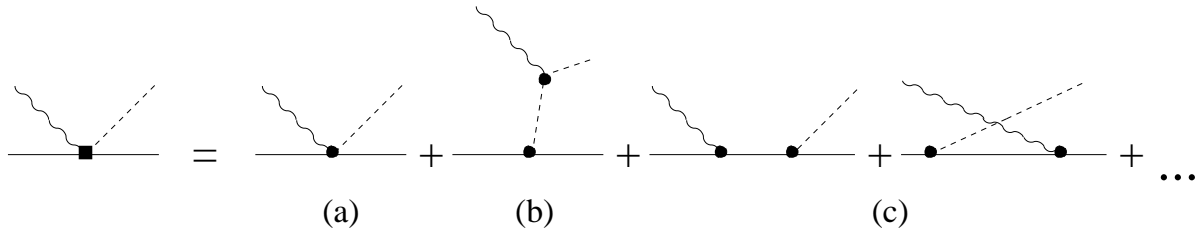, height=3cm, angle=0}
\caption{Diagrams that contribute to the one--body production operator in
  diagrams (a1) and (a2) of Fig. \protect{\ref{diagam}}. Only
tree level diagrams that contribute to $\gamma p\to \pi^+n$ up to
  order $\chi^2$ are shown. The dots denote loops that contibute to the
  $s$--wave amplitude at oder $\chi^2$ as those shown in  Fig. \protect{\ref{piinfl}} (b)
  and (c).}
\label{pwaves}
\end{center}
\end{figure}

We will start with a discussion of the one--body contributions to be used in
the diagrams of type (a1) and (a2) of Fig.~\ref{diagam}.

At threshold
the $\pi \gamma NN$ vertex at leading order ($\chi_m^0$)
 and at next--to--leading order ($\chi_m^1$) is given
by the so--called
Kroll--Ruderman (KR) term~\cite{KrRu} and its recoil correction
\footnote{A list of the vertices relevant for chiral perturbation theory
  calculations can be found in Ref.~\cite{ulfs}; however, in contrast to that
  reference we use the spinor normalization $\bar u u = 2M_N$. In addition we
  used the Goldberger--Treiman relation to replace $g_A/f_\pi$ by $g_{\pi
  N}/M_N$.},
\begin{equation}
\hat V_{\pi\gamma NN}^{KR}
=ieg_{\pi N}\left(1-\frac{\omega_\pi}{2M_N}\right)
\left(\vec \epsilon_{\gamma} \cdot \vec \sigma\right) \epsilon^{3ab} \tau^b \ ,
\label{kr}
\end{equation}
where $\vec \epsilon_{\gamma}$ denotes the photon polarization and
$\omega_\pi$ is  the energy of the outgoing pion.  The
corresponding diagram is shown as diagram~(a) in Fig.~\ref{pwaves}.
  Note that we use $g_{\pi N}=13.4$ in the calculation and
  the charge $e$ is normalized such that the fine structure constant is given
  as $e^2/(4\pi)=\alpha=1/137.$
 This vertex 
contributes to the one--nucleon operator (impulse term) as shown in diagram
(a1) and (a2) of Fig.~\ref{diagam} and also provides the production vertex for
the virtual pion to be rescattered.  As the $\pi \gamma NN$ vertex is both
spin and isospin dependent, now, in contrast to $\pi d$ scattering, for the
rescattering contributions the two--nucleon intermediate state that occurs
while the pion is in flight is allowed by the Pauli principle to be in an $S$--wave.

According to table \ref{Table_1}
at order $\chi$ we also need to consider the leading correction in
$\chi_Q^2$ to the $s$--wave as well as the leading $p$--wave
contribution which is of order of $\chi_Q$.
Both are provided by the diagram where the photon couples to the pion in
flight  
corresponding to  diagram~(b) of Fig.~\ref{pwaves}.
The corresponding expression for the effective $\pi \gamma NN$ vertex
reads
\beqa
\hat V_{\pi\gamma NN}^{(b)}
={-ieg_{\pi N}}\vec \sigma \cdot \left(\vec k_\pi-\vec q_\gamma\right)
\frac{\vec \epsilon_{\gamma} \cdot \vec k_\pi}{\omega_\pi
 |\vec q_\gamma\, |-\vec k_\pi \cdot
  \vec q_\gamma} \epsilon^{3ab} \tau^b \ ,
\label{vpiif}
\eeqa
where $\omega_\pi$ denotes the energy of the outgoing pion.
This vertex contributes to the leading p--waves (order
$\chi_Q$ for the amplitude --- order $\chi_Q^2=\chi$ for the total cross section),
the leading energy dependence of the $s$--wave (order $\chi_Q^2$
for the amplitude as well as the cross section, for this term can interfere
with the contribution of order $\chi^0$),
as well as the leading $d$--waves (order $\chi_Q^2$ for the amplitude and thus
order $\chi_Q^4=\chi^2$ for the cross section).

\begin{figure}[t]
\begin{center}
\epsfig{file=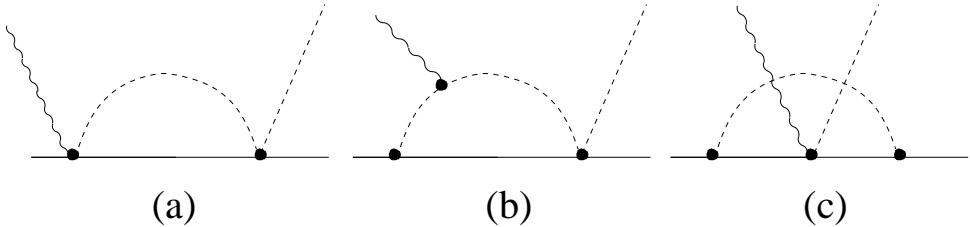, height=3cm, angle=0}
\caption{Typical pion loops that contribute to $\gamma p\to
  \pi^+n$. Crossed terms for diagrams (a) and (b) are not shown explicitly.}
\label{piinfl}
\end{center}
\end{figure}

At order $\chi_m^2$ pion loops start to contribute to the $s$--wave
part of the $\gamma p\to \pi^+ n$ amplitude (see
Fig.~\ref{piinfl}) and thus to the reaction $\gamma d\to \pi^+ nn$ as well.
The only additional vertex needed for the evaluation of these loops is given in Eq. (\ref{pinpot}) with
$g_-=-\sqrt{2}m_\pi/(2f_\pi^2)$
and $g_+=0$.
 The imaginary part of  diagram~(a) in
Fig.~\ref{piinfl} contributes at  $\chi_m^2 \chi_Q$ whereas 
the imaginary part of diagram~(b)  starts to contribute only at higher order
($\chi_m^2 \chi_Q^3$) and is thus not explicitly included in the calculation.
In contrast to the imaginary parts the real parts of 
some of those  pion loops on a single nucleon 
are divergent and need to be regularized (e.g. diagram~(c) of Fig.~\ref{piinfl})
\footnote{For a complete discussion of all relevant loops we refer  to Refs. \cite{threspiprod,pi0photo}.}.
In the  calculation of the  reaction $\gamma d\to \pi^+ nn$ 
  we use a prescription
similar to that used for $\pi d$ scattering, as described in the previous
section. In practice this means to replace the expression of Eq.~(\ref{kr}) by
\begin{eqnarray}
\hat V_{eff}= 
i\kappa E_{0+} (\vec\epsilon_{\gamma} \cdot \vec \sigma) \epsilon^{3ab}
\tau^b \ ,
\label{veff}
\end{eqnarray}
where $\kappa = 4\pi\sqrt{2}\left(m_\pi+M_N\right)$.
The experimental value of $E_{0+}$ 
 from the reaction $\gamma p\to \pi^+ n$ is \cite{saskat}: 
$$E_{0+}=(28.06\pm0.27\pm 0.45)\times 10^{-3} m_\pi^{-1}.$$
This value  coincides within errors
 with the result of ChPT up to order $\chi_m^3$ \cite{BerKai}:
\beq E_{0+} = (28.2\pm0.6)\times 10^{-3}m_{\pi}^{-1}
\label{E0thr} .\eeq
Although this calculation was performed to higher order than what we aim at
here, we will use the latter value as input quantity for the threshold value
of $E_{0+}$ since the contribution of order $\chi_m^3$ are of the order of the
assigned uncertainty.

It is straightforward to show that the $s$--wave contribution
derived from Eq.~(\ref{vpiif}) 
 has the same 
operator structure as  Eq.~(\ref{veff}) and we may simply include its effect 
 by replacing $E_{0+}$ in Eq.~(\ref{veff}) by $E_{0+}(k_\pi^2)$
defined up to order $\chi_Q^4$ through
\begin{eqnarray}
E_{0+}(k_\pi^2)= E_{0+}+ E_{0+}^{\, \prime}
 \frac{k_\pi^2}{m_\pi^2}
+ E_{0+}^{\, \prime \prime}
 \frac{k_\pi^4}{m_\pi^4} \ .
\label{eedep}
\end{eqnarray}
The leading contribution to both coefficients  $E_{0+}^{\, \prime}$ and $E_{0+}^{\, \prime\prime}$ can be
calculated directly from Eq.~(\ref{vpiif}). At next to leading order the
energy dependence of the $\gamma \pi NN$ vertex enters as well (c.f. Eq.~(\ref{kr})). Especially, we find
$$
E_{0+}^{\, \prime}=-\frac{eg_{\pi N}}{3 \kappa}\left(1+\frac{m_\pi}{M_N}\right) =
 -11.3  \times 10^{-3}m_{\pi}^{-1} \ ,
 $$
 where $|\vec q_\gamma|=m_\pi(1-m_\pi/(8M_N)+O(\chi^2))$ was evaluated in
 the center of mass system of the reaction $\gamma d\to \pi^+ nn$ at threshold.
The slope parameter $E_{0+}^{\, \prime}$ is calculated up to its leading and next--to--leading
order. However, we anyhow assign an uncertainty of 10 \% to this quantity, for
higher order corrections are known to be enhanced by the $\Delta$
resonance \cite{heavyChPT}.
The given value for  $E_{0+}^{\, \prime}$ compares well to that
 from the dispersive analysis of the
Mainz group \cite{disp}
and is consistent 
with that used as input in Ref.~\cite{saskat} to extract
$E_{0+}$ from the data.

At order $\chi^2$ for our reaction, additional diagrams contribute to the transition $\gamma p\to \pi^+n$
where the photon gets absorbed on the nucleon, e.g. through the 
magnetic moment, followed by pion emission (diagrams (c) in Fig.~\ref{pwaves}).
Thus, they are accompanied by a one--nucleon intermediate state. These
diagrams, when used as the one--body operator in diagram~(a2) of Fig.
\ref{diagam}, acquire a two--nucleon cut that leads to an imaginary amplitude
even at threshold due to the kinematically allowed transition $\gamma d\to pn$
followed by $pn\to \pi^+ nn$. 
However, this two--nucleon cut introduces a new (large) momentum scale
$p\simeq \sqrt{m_\pi M_N}$ into the problem that calls for special care.
According to the counting rules for pion production in $NN$
collisions~\cite{pionprod}, 
the contribution of this two--nucleon cut 
is therefore suppressed by $\chi_m^{3/2}$ compared to the rescattering diagram
(c2) of Fig.~\ref{diagam}.  
  It is therefore justified within the ordering scheme
used to replace the two--nucleon propagator by its static limit.
The resulting contribution from the magnetic couplings to the $s$--wave pion production on the nucleon
is already
included in the effective operator of Eq.~(\ref{veff}).
For the pion $p$--waves we get the following expression for the sum of the
$s$-- and $u$--channel contributions  of Fig.~\ref{pwaves}(c) 
\beqa
\hat V_{\pi\gamma NN}^{(c)}
=-\frac{ieg_{\pi N}}{\sqrt{2}M_n m_\pi}\left(A+\vec \sigma \cdot \vec B
\right) \ ,
\label{magmom}
\eeqa where \beqa A&=&i(\mu_p-\mu_n)\vec k_\pi \cdot (\vec \epsilon_\gamma
\times \vec q_\gamma)
\ , \\
\vec B&=&-2\left(\vec \epsilon_\gamma\cdot \vec p\right)\vec k_\pi
+(\mu_p+\mu_n)\left((\vec \epsilon_\gamma\cdot \vec k_\pi)\vec q_\gamma -
  (\vec q_\gamma\cdot \vec k_\pi)\vec \epsilon_\gamma\right) \ . 
\label{vecB}
\eeqa Here $\vec
p$ denotes the momentum of the incoming nucleon and $\mu_p=2.79$ and
$\mu_n=-1.91$ denote the magnetic moments of the proton and the
neutron, respectively.  
The operator  $\hat V_{\pi\gamma NN}^{(c)}$ 
given by Eq.~(\ref{magmom}) was evaluated  directly for the channel  $\gamma p\to \pi^+ n$.
As a consequence the
expression does not show any explicit isospin dependence and an isospin factor
of $\sqrt{2}$ appears. 

In Ref.~\cite{heavyChPT} it is shown that the pion $p$--waves converge quite
slowly: the next--to--leading order correction to the dominant $p$--wave
multipoles $M_{1^+}$ and $M_{1^-}$ change the leading result by a factor of 2.
The reason for this sizable correction is the large numerical value of
$(\mu_p-\mu_n)=4.7$. On the other hand, the contribution of the $A$--term to
$\gamma d\to \pi^+ nn$ is suppressed. One reason can be read off from
Eq.~(\ref{magmom}) almost directly: since the $A$--term is a scalar in spin
space, it will not change the total spin of the two nucleon system, when used
as the one--body operator in diagram~(a) of Fig.~\ref{diagam}. Thus the final
$nn$ state is in a spin--triplet state. However, in the near threshold regime
of interest here, the simultaneous appearance of an $nn$ $P$--wave ($nn$
spin--triplet states are to have odd angular momenta as a consequence of the
Pauli principle) and a pion $p$--wave is suppressed. In addition, it turns out
that for the total cross section the $A$--term does not interfere with the
leading pion $p$--wave contributions (the term proportional to $\vec q_\gamma$
in Eq.~(\ref{vpiif})) which leads to an additional suppression. For details on
the latter point we refer to the explicit expressions given in appendix A.
One should also note that like for  the slope of the $s$--wave
amplitude, the $\Delta$--isobar gives potentially large corrections to the $p$--wave amplitudes \cite{heavyChPT}. 
Thus we assign an uncertainty of 10\% also to those.

At order $\chi^2$, pion rescattering diagrams (see diagrams (b), (c) and
(d) in Fig.~\ref{diagam}) start to contribute.  Here the same
diagrams contribute as previously discussed for $\pi d$ scattering, but with
the first $\pi N$ interaction being replaced by the $\gamma N\to \pi N$
transition vertex.  These diagrams are depicted in Fig.~\ref{diagam} (b2) and
(c2). However, there is in addition a whole new class of diagrams, namely those
without final state interaction (depicted as (b1) and (c1)). Another important
difference to $\pi d$ scattering is that, as already mentioned above, the S-wave
two--nucleon state---while the pion is in flight---is now allowed by the Pauli
principle.  This has two consequences: first of all we expect that the static
approximation will not work for the rescattering contribution and, as a second
consequence, now nothing prevents the two--nucleon system to interact while
the pion is in flight. As the $NN$ interaction is to be taken into account to
all orders, these diagrams are also potentially important.  The latter
statement becomes especially clear when observing that the scattering length
in the $nn$ channel is quite large; note that in the limit of an infinite
scattering length the diagrams (d1) and (d2) acquire a logarithmic infrared
divergence. In fact, then the two--nucleon propagator behaves as if it would
describe the propagation of a massless particle\footnote{In a different
  scheme and for a different reaction this behavior was studied in Ref.~\cite{martin}.}.

Starting from the vertices given by  Eq.~(\ref{kr}) and in the appendix of Ref. \cite{ulfs} 
it is straightforward to write down the corresponding matrix elements for the
diagrams shown in Fig.~\ref{diagam}. Note that all diagrams are
evaluated in Coulomb gauge. This is a standard choice in ChPT calculations,
for many diagrams with single photons are relegated to higher orders, such as
those where the photon couples to the  charge of the nucleon and of the
deuteron, respectively.

\begin{figure}[t]
\begin{center}
\epsfig{file=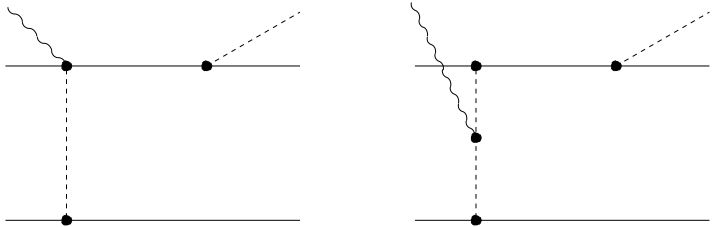, height=3cm, angle=0}
\caption{Diagrams where the photon couples to the structure of the
  deuteron. These diagrams start to contribute only at order $\chi^3$ and are
  therefore not considered here.}
\label{structure}
\end{center}
\end{figure}

Diagrams, where the photon couples to a rescattered pion in flight turn out to
be strongly suppressed numerically compared to those shown in
Fig.~\ref{diagam}.  The reason for this is twofold: their imaginary part is
suppressed as a consequence of gauge invariance that forces the photon--pion
coupling to be of the order of the (small) pion momentum in the loop and the
appearance of a second pion propagator leads to an additional suppression.
This is in complete analogy to $\pi d$ scattering as discussed in detail in
Ref.~\cite{pid}.  In addition gauge invariance in principle also calls for the
inclusion of diagrams, where the photon couples to the internal structure of
the deuteron. Typical representatives of this class are shown in figure
\ref{structure}. However, it is easy to show that these diagrams are
suppressed by at least one power of $\chi$ compared to the leading
rescattering contribution and therefore start to contribute at most at order
$\chi^3$ to $\gamma d\to \pi^+nn$\footnote{For this estimate we used $m_\pi$
  as typical momentum in the deuteron. A more accurate estimate of this
  quantity would be $\gamma=\sqrt{E_BM_N}$. If we were to use this for the
  power counting it would yield a suppression of the order of $\gamma/M_N \sim
  0.3 \chi$.}.  Therefore these diagrams are not included in the calculation.
As a consequence, the diagrams given in Fig.~\ref{diagam} (together with
Figs.~\ref{pwaves} and \ref{piinfl}) are all that contribute to the given
order.

%
%

Details on how the various diagrams are evaluated are given in the
appendix A. Especially, there it is explained how we dealt with the
three--body singularities that occur in
 diagrams (c2), (b2),  (d1), and (d2).
We want to mention that we evaluated the loop diagrams for non--relativistic 
pions (see appendix A) for that largely simplified the numerics. We checked by direct
evaluation of diagrams (c1) and (c2) for threshold kinematics  that switching to relativistic pions changes the
individual contributions to the amplitude by less than 4 \%.

\begin{figure}[t]
\begin{center}
\epsfig{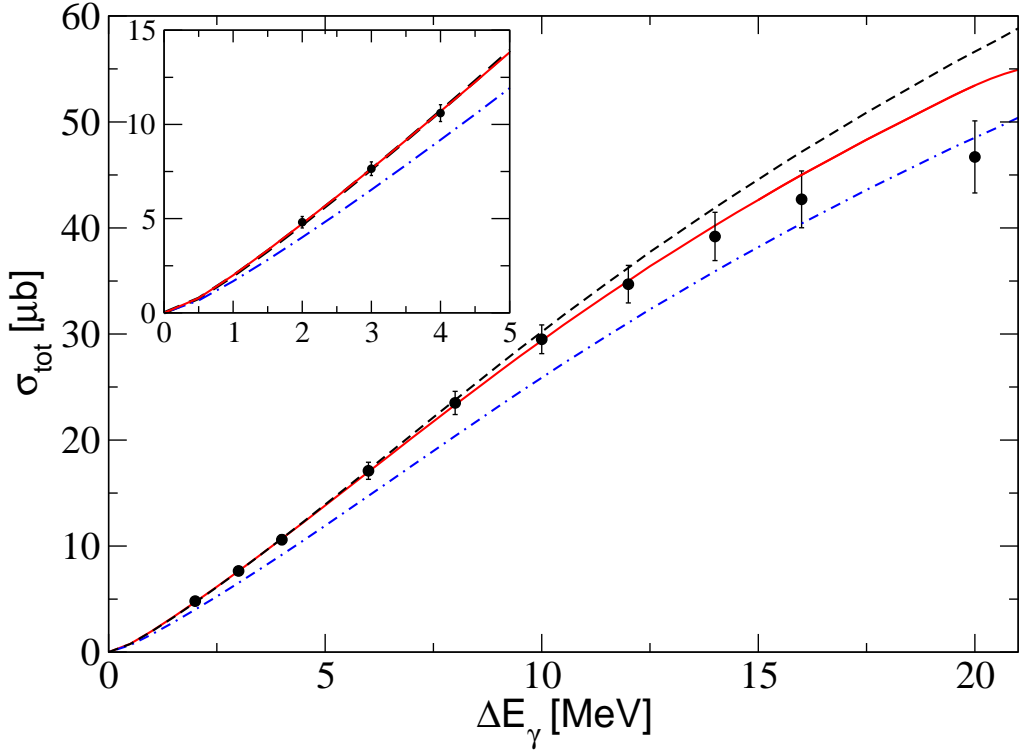}
\caption{ Total cross section of the reaction $\gamma d\to \pi^+ nn$
at LO (dashed line), NLO (dash--dotted line) and $\chi^{5/2}$ order (solid line)
 together with experimental data from Ref.~\cite{MIT}.
}
\label{results}
\end{center}
\end{figure}

\section{Results and discussion}
\label{resadis}

In our calculations we use the standard values for the various constants,
namely $f_\pi=92.4$ MeV, $m_\pi=139.57$ MeV (only the charged pions contribute
to the order we are working) and $M_N = 938.27$ MeV.
The deuteron wave function and also the scattering wave
functions needed for the $nn$ scattering amplitudes (in the
$^1S_0$, $^3P_0$, $^3P_1$, and $^3P_2$ partial waves) 
are generated from the (charge--dependent) CD-Bonn potential \cite{CDBonn}.
Specifically, for the former we employ the analytical
representation of the deuteron wave function provided in that reference
because it allows us to perform some integrations analytically---see appendix B.
For the same reason the scattering wave functions are computed from rank-1 
separable representations of the CD-Bonn model, constructed along the lines of 
Ref.~\cite{Johann} utilizing the so-called Ernst--Shakin--Thaler method \cite{Johann2}, cf. also appendix B.  
Note that the scattering length predicted by the CD-Bonn potential for 
the $^1S_0$ partial wave in the $nn$ system is $a_{nn} = -18.97 $ fm \cite{CDBonn},
which is in line with most of the recent experimental information \cite{Howell,Trotter}
(note, the analysis of Ref. \cite{bn} gave a significantly lower value).
We want to point out that in our calculation the contribution of the deuteron 
D-wave is included. This contribution to the leading diagrams is rather important,
 because it guarantees the correct normalization of the S-wave component for the potential used.

Although it would be desirable to use wave functions consistent with the
 transition operator, as provided, e.g., in Ref.~\cite{evgeni}, various
 studies show a low sensitivity to the wave function employed (see, e.g., \cite{pid}).
In addition, the current operators have not yet been worked out to the same
 order and within the same unitary transformation as the $NN$ wave functions.
We postpone this to a future study.

\begin{figure}[t]
\begin{center}
\epsfig{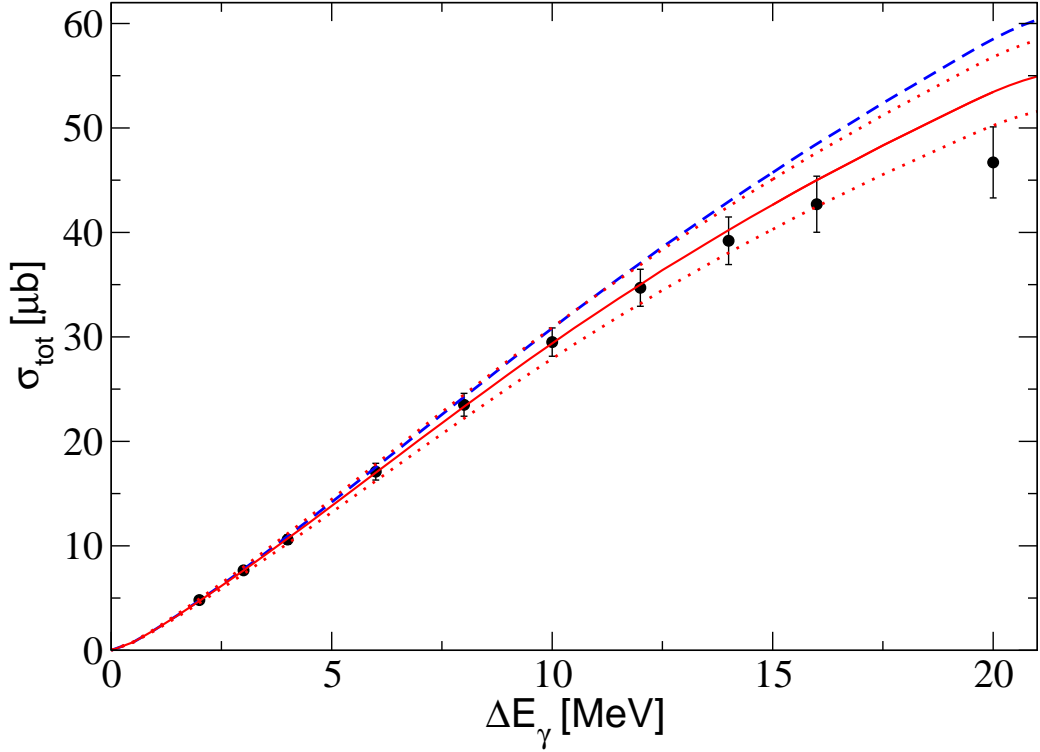}
\caption{Importance of the slope parameter $E_{0+}^{\, \prime}$ defined in
  Eq. (\protect\ref{eedep}):
The solid line shows the full result,
 whereas the long dashed line is produced with  $E_{0+}^{\, \prime}=0$.
The two dotted lines show the estimated uncertainty of  full result as
 described in the text. 
 The
 data are from Ref.~\cite{MIT}.
}
\label{slope}
\end{center}
\end{figure}

The reaction $\gamma d\to \pi^+ nn$ was calculated in the DWBA approximation
by many authors in the middle of the seventies (cf., e.g., the review paper by
Laget \cite{Lagrev} and references therein).  The approach used in those
investigations corresponds to the evaluation of diagrams (a1) and (a2) of Fig.~\ref{diagam}
using the $\gamma p\to \pi^+ n$ vertex of Eq.~(\ref{kr}); thus, in our
language those were incomplete calculations up to next--to--leading order
($\chi$), because the energy dependence of the $\gamma \pi NN$ vertex and
higher pion partial waves were neglected.
Since the most
important contribution to the $\gamma p\to \pi^+ n$ operator in the near
threshold region originates from the Kroll-Rudermann operator and its first
correction (see Eq.~(\ref{kr})), which is known, and the convergence of
diagram~(a2) of Fig.~\ref{diagam} is provided by the universal fall-off of the deuteron 
wave function for small momenta (fixed by the deuteron binding energy) 
all those calculations
led to very similar results (that we reproduce).  To improve the calculations,
some authors used instead of the prefactors of Eq.~(\ref{kr}) the experimental
input for $E_{0+}(k_\pi^2)$ at threshold as the strength parameter for the one--body term (see,
e.g., Ref.~\cite{MIT}). This, however, corresponds to an
incomplete order-$\chi^2$ calculation, as discussed above. In addition, none of the works reported in 
Ref.~\cite{Lagrev} considered the $D$--wave of the deuteron consistently. The slope of $E_{0+}(k_\pi^2)$,
pion $p$--waves, or the $NN$ final state interaction in the $P$--waves were also not considered.  The
only attempt to improve the mentioned calculations via inclusion of a pion
rescattering contribution was made in Ref.~\cite{Tzara}, where diagram~(c2) of
Fig.~\ref{diagam} was evaluated in the static limit, i.e. without nucleon recoil.  This
contribution was found to be large, amounting to an increase of around 10\% of
the total cross section (see the discussion of this question in 
Ref.~\cite{Lagrev}). However, as we stressed above, the static approximation is
expected to work very poorly in this reaction. 

\begin{figure}[t]
\begin{center}
\epsfig{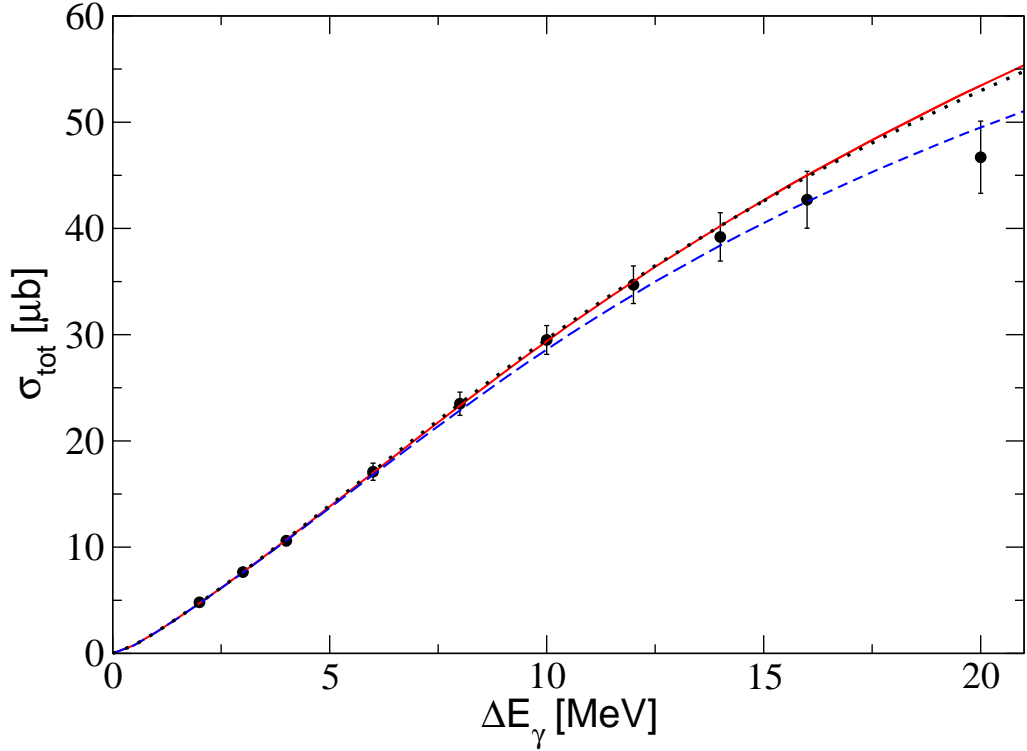}
\caption{The role of the higher pion partial waves and effect of the $NN$
  interaction in the
  $P$--wave.
The   dashed line shows the calculation where only the S-wave
interactions are taken into account,  the dotted
  line includes also the higher pion partial waves  and the solid line
includes in addition the contribution of the $NN$ final state interaction in the
 three possible  included in diagram~(a2) of Fig.~\ref{diagam}.}
\label{Pwaves2}
\end{center}
\end{figure}

Our results at order $\chi^0$, $\chi$ and  combined order $\chi^2$ and $\chi^{5/2}$
 are shown in Fig.~\ref{results} by the  dashed, dash--dotted and solid
lines, respectively.
The data are taken from Ref.~\cite{MIT} and the energy is measured in terms of
 $\Delta E_\gamma= (1{+}m_\pi/(2M_N))Q$---the photon lab energy subtracted by
 its threshold value.
Up to order $\chi^1$ the result is parameter free, i.e. it is
 determined by the value of  $g_{\pi N}$ only.
 For the value of $E_{0+}(k_\pi^2)$ at threshold that is needed as the only input quantity at
order $\chi^2$, we use the central values of the ChPT calculation of Ref.~\cite{BerKai}
(that agrees with the experimental number,
as mentioned in the previous section).

\begin{figure}[t]
\begin{center}
\epsfig{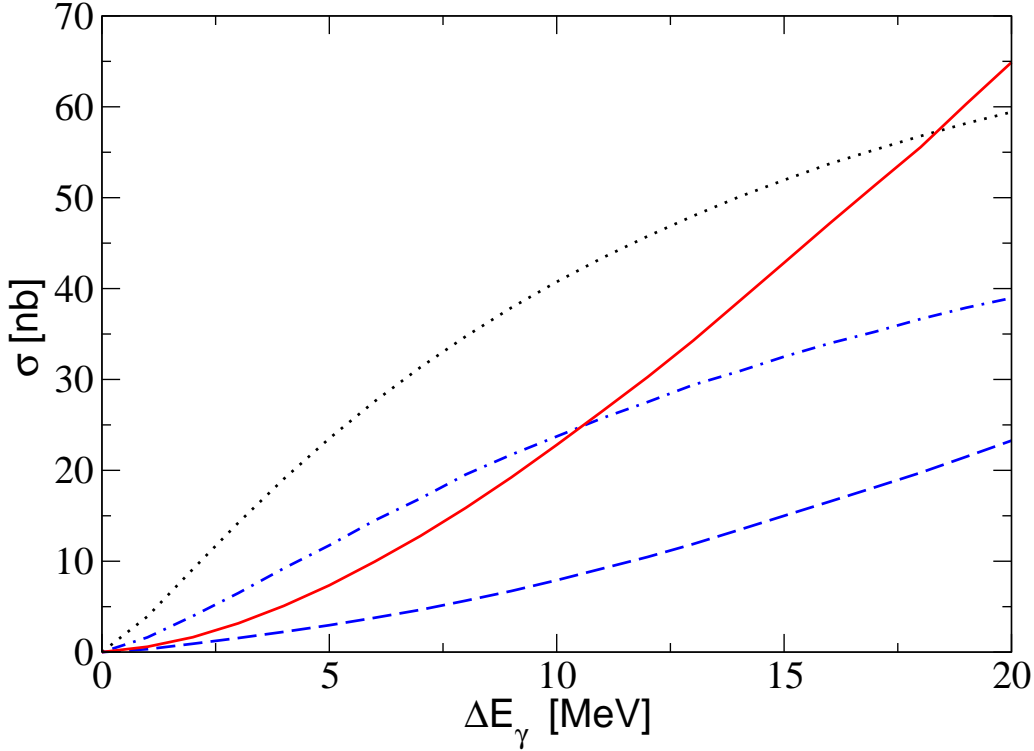}
\caption{Quality of the static approximation for diagram~(c2) of Fig.~\ref{diagam}: The
dash--dotted line shows the full calculation for diagram~(c2), the dashed
  one that for diagram~(b2), and the solid line the sum of the two. On the
  other hand, the result for diagram~(c2) evaluated in the static approximation
  is given by the dotted line.}
\label{stvsfull}
\end{center}
\end{figure}

To illustrate the importance of the slope  $E_{0+}^{\, \prime}$ we show in
Fig.~\ref{slope} both the full result as well as the one we get
for a vanishing slope. For $\Delta E_\gamma$ larger than 10 MeV we observe a quite
large sensitivity to the energy dependence of  $E_{0+}(k_\pi^2)$.

The counting scheme of Eq.~(\ref{expansion}) calls for an inclusion of
$p$-- and $d$--waves for the pion. The explicit expressions for the
vertices are given in Eqs.~(\ref{vpiif}) and (\ref{magmom}). These vertices are to be considered embedded in
diagrams (a1) and (a2) for Fig.~\ref{diagam} only. The contribution of the
higher pion partial waves can be read off the difference of the dashed and the
dotted line in Fig.~\ref{Pwaves2}.

The significance of $NN$ $P$--waves, even for $\Delta E_\gamma \le 20$ MeV,
was established long ago \cite{Lagrev}, however, so far they were only
included as plane waves (through diagram~(a1) of Fig.~\ref{diagam}).  In this
work for the first time the final state interaction in the $NN$ $P$--waves is
considered as well: we include the $NN$ interaction in the three possible
$P$--waves ($^3P_0, ^3P_1$ and $^3P_2$) in diagram $(a2)$.  The effect of the
nucleon $P$--waves on the total cross section is shown in Fig.~\ref{Pwaves2}.
It is interesting to note that although the resulting effect of the $P$-wave
$NN$ interaction is not significant for the total cross section the individual
contributions of different partial waves are quite large (e.g., the $^3P_1$
partial wave enhances the cross section by about 15\% at $\Delta
E_{\gamma}=20$ MeV). It turns out that the contribution of the $^3P_1$ partial
wave, which is enhanced due to constructive interference of the diagram~(a1)
and (a2) of Fig.~\ref{diagam} for this particular $NN$ partial wave, cancels
to a large extent the contributions of the $^3P_0$ and $^3P_2$ partial waves.
Note that the interference pattern of the various $NN$ $P$--waves can be
traced to the fact that $^3P_1$ is repulsive, whereas $^3P_0$ and $^3P_2$ are
attractive for small relative momenta. We have to mention that in the present
calculation we neglect the coupling of the $^3P_2$ partial wave to the
$^3F_2$. However, in view of the cancellation mentioned above this coupling
might play a role, specifically because the mixing parameter $\epsilon_2$ is
already about one third of the corresponding $^3P_2$ phase shift at $NN$
energies corresponding to $Q=20$ MeV. Also the $NN$ $^1D_2$ partial wave,
which is likewise not included in the present calculation, should start to  play a
role with increasing energy.  As mentioned above, the $NN$ wavefunctions used are
not consistent with the transition operator. As long as this is not the case
we do not think that including these probably minor effects is worth the
effort. However, once wave functions consistent with the chiral counting are
used, also the mentioned approximations need to be abandoned.

To summarize the results, we have calculated higher order chiral corrections
to the impulse term used, e.g., in Ref.~\cite{Lagrev}. We found that due to a
compensation amongst the effects of the various $NN$ $P$--waves, their net
effect on the total cross section is negligible. In addition, the effect of
the energy dependence of the pion $s$--wave production (parametrized by the
slope $E_{0+}'$) and of the pion $p$--waves compensated each other largerly as
well.  In view of this findings, the good agreement of the results of, e.g.,
Ref.~\cite{Lagrev} for the total cross section for photon energies above 10
MeV with the data should be considered accidental.  However, for differential
observables and especially for polarization observables we expect sizable
effects from the chiral corrections calculated in this work.

%
%

Let us now discuss some of the results in detail.
We found that all rescattering contributions of order $\chi^2$ and of order $\chi^{5/2}$ 
  contribute with similar strength. This is true also for 
those diagrams where the $NN$ interaction appears inside the loop (diagrams
(d1) and (d2) of Fig.~\ref{diagam}).
This clearly demonstrates the need to take into account
  the $NN$ $S$--wave interaction non--perturbatively also in intermediate states (as it
  is done routinely in three-nucleon calculations anyway). 
  
  Let us now compare in detail the results of different diagrams at the
  threshold.  Here contributions from higher partial waves as well as those from the
  slope parameter $E_{0+}'$ vanish.  Therefore the uncertainty of the
  calculation in this regime can be estimated much more accurately than at
  higher energies.  The results of the calculation of all possible
  contributions to the reaction amplitude at threshold are given in table
  \ref{contrib}. Evidently, the total contribution from few--body
  (rescattering) corrections to the full amplitude at threshold is about 5~\%
  of the contribution from the distorted wave impulse approximation.  One can
  also see from this table that the net contribution of the orders $\chi_m^2$
  and $\chi_m^{5/2}$ cancels largely the NLO contribution (see also the
  corresponding results in Fig.~\ref{results} at low energies). Due to this
  one might speculate that corrections to this result from
  higher orders, especially from $\chi_m^3$, may influence the results
  stronger than suggested by the power counting.  However
  we think that the uncertainty of the calculation at low energies is indeed
  within the uncertainty for $E_{0+}$ given by Eq.~(\ref{E0thr}).  Actually, no
  additional diagrams to those shown in Fig.~\ref{diagam} appear at order
  $\chi_m^3$. Thus, the contributions to the amplitude at this order originate
  basically from three sources:
\begin{enumerate}
\item NLO correction to the Kroll-Rudermann vertex when being used as a
  one--body operator in diagrams (b),(c) and (d) of Fig.~\ref{diagam} and the
  recoil corrections to the Weinberg--Tomozawa term. These corrections are 
  of order $(m_{\pi}/2M_N)\cdot 5\%=0.4\%$.
\item Relativistic corrections to the diagrams with pion rescattering. For
  diagram~(c2) the relativistic correction gives only about 4\% (to the 3 \%
  contribution to the amplitude from diagram (c2)). Since there are no reasons a priori which could enhance the
  corresponding corrections to the other rescattering diagrams, we assign an
  uncertainty of order of 0.5 \% to this effect.
\item The contributions from the coupling to the deuteron structure and the
  $nn$ scattering $T$--matrix. These are
  the hardest to estimate. However, as argued above, they are expected to be
  numerically quite small (see footnote \# 6).
\end{enumerate}
To summarize, we assign an
uncertainty of 2 \% to the few--body corrections of the amplitude.
On the other hand
the contribution at order $\chi_m^3$ to  $E_{0+}$ at threshold is also
  about 2 \% \cite{BerKai}. Adding these two uncertainties in quadrature we
  end up with a total uncertainty of 3 \% for the transition operator near threshold.
Unfortunately we are not in the position to estimate the uncertainty from the
$NN$ interaction used---this has to be postponed until a calculation with
fully consistent wave functions is performed. 

In Fig.~\ref{slope} the resulting uncertainty for the full calculation is
shown by the two dotted lines. This range of uncertainties contains besides
the 3~\% just discussed also the 10 \% uncertainties on both the pion
$p$--waves as well as the slope parameter.

\begin{table}[t]
\renewcommand{\baselinestretch}{1}
\begin{center}
\vskip 0.2cm
{\large
\begin{tabular}{|c|c|c|c|}
\hline
operator & order  & diagrams & contribution\\
\hline
&  & & \\
 & $\chi_m^0$ & (a1)+(a2) & 1 \\
& & & \\
one-body         & $\chi_m^1$ & (a1)+(a2) & $-$0.07\\
& & & \\
         & $\chi_m^2$ & (a1)+(a2) & $+$0.028 \\
& & & \\
\hline
& & & \\
         & $\chi_m^{5/2}$& (b2) & $-$0.016\\
& & & \\
         & $\chi_m^2$,$\chi_m^{5/2}$ & (c1)+(c2) & $+$0.039\\
few-body& & & \\
         & $\chi_m^2$,$\chi_m^{5/2}$ & (d1) & $+$0.008\\
& & & \\
         & $\chi_m^2$,$\chi_m^{5/2}$ & (d2) & $+$0.024\\
& & & \\
\hline
\end{tabular}}
\caption{Contributions of different diagrams and operators to the reaction amplitude at threshold are given 
at different orders. 
The results in table are normalized to that of the calculation of diagrams 
(a1) + (a2) with the leading order Kroll-Rudermann vertex. Note that diagram
(b1) does not contribute at threshold.}
\label{contrib}
\end{center}
\end{table} 

To show that the static exchange is indeed a poor approximation to the exact
result for diagram~(c2) of Fig.~\ref{diagam}, as conjectured above, we compare in Fig.
\ref{stvsfull} the results for the static approximation with those from the
exact calculation as well as with those for the one--body term (b2). As can be
clearly seen, the static approximation fails to describe the full result in
both strength as well as energy dependence. Evidently, for
the exact calculation the total contribution of the sum of the results for (b2)
and (c2) is extremely small near threshold, i.e.  in the region where the real
parts of both diagrams dominate. With increasing energy the contribution from
the sum of these two diagrams increases rapidly. The reason for this effect is
that at higher energies the role of the imaginary parts of both diagrams,
which contribute coherently for the Pauli allowed $NN$ states, is growing
rapidly.  The resulting contribution for the sum of both diagrams of Figs.
2 (b2) and 2 (c2) to the total cross section does not exceed 4\%.

\section{Summary}

In this paper we presented for the first time a ChPT calculation for
the reaction $\gamma d\to \pi^+ nn$. 
We calculated the diagrams displayed in Fig.~\ref{diagam}---keeping explicitly 
the nucleon--recoil for the intermediate states. This corresponds to a complete
calculation up to order $\chi^{5/2}$, where $\chi=m_\pi/M_N$.

The results of the full calculation are shown in Fig.~\ref{results} by the
solid line. A good agreement between theory and experiment at low energies is obtained without
any free parameter.  The only input parameter was the threshold value for
$E_{0+}$, taken from an N$^3$LO calculation for $\gamma p\to \pi^+
n$ \cite{BerKai}. Estimated conservatively, the uncertainty  for the
transition operator of our calculation is about 3 \% in the
amplitude for photon energies $\Delta E_\gamma$ below 5 MeV.
Unfortunately we are not in the position to estimate the uncertainty from the
$NN$ interaction used---this has to be postponed until a calculation with
fully consistent wave functions is performed. 


We found a strong suppression of the pionic 
rescattering contributions
in comparison to the  calculation in the frozen-nucleon 
approximation, i.e. in the static limit. This confirms 
the suggestion made in
Ref.~\cite{Lagrev} and is in line with the general remark of 
our recent paper \cite{rec} where it was conjectured that 
the static limit is not adequate for pion rescattering processes with  Pauli--allowed 
S-wave intermediate $NN$ states.

As a next step a fully consistent chiral perturbation theory
calculation for the given reaction should be performed. 
It should include not only the use of
wave functions constructed within this framework, as provided in Ref.~\cite{evgeni}, 
but also a re-evaluation of the transition operator within the
same unitary transformation used to calculate the wave functions. Only then
one
can 
reliably asign a theoretical uncertainty to the full calculation and 
address questions raised in Ref.~\cite{rus} for the consistency of the
counting scheme and the scaling of four--nucleon operators. Although we do not
expect the latter effects to be numerically significant for the reaction considered
here, they are potentially important for $\gamma d\to \pi^0pn$, where the
leading contribution vanishes.

Once a fully consistent ChPT calculation and better data are available for the reaction 
$\gamma d\to \pi^+ nn$ one can consider to extract the energy dependence of 
$E_{0+}$ from this reaction, for 
the total cross section is very sensitive to the slope of $E_{0+}$, as illustrated in
Fig.~\ref{slope}.
However, before more definite conclusions can be drawn on this issue better
data on the total cross section as well as the differential cross section
in the energy regime of 15--25 MeV is needed.

The work presented allows one, amongst other things, to address the
theoretical uncertainty of the $nn$ scattering length extracted from $\gamma
d\to \pi^+ nn$ analogously to the studies of Ref. \cite{daniel} for $\pi^-d\to
\gamma nn$. These studies are of high interest in the light of the significant
differences in the values for $a_{nn}$ extracted from different
groups in different reactions---c.f., e.g., Refs. \cite{Howell,Trotter} and Ref. \cite{bn}.

\vspace{1cm}

\noindent 
{\bf Acknowledgment}

\noindent 
We are 
thankful to A.Gasparyan and L. Tiator for useful discussions.
We also thank D.R. Phillips for a stimulating discussion and useful comments
to the manuscript.
This research is part of the EU Integrated Infrastructure Initiative
Hadron Physics Project under contract number RII3-CT-2004-506078, and was 
supported also by the DFG-RFBR grant no. 02-02-04001 (436 RUS 113/652)
and the DFG--Transregionaler--Sonderforschungsbereich SFB/TR 16 "Subnuclear Structure of Matter".
A.E.K, and V.B. acknowledge the support of  
the Federal Programme of the Russian Ministry of Industry, Science, and Technology No 40.052.1.1.1112.

\appendix

\newpage

\begin{figure}[t]
\begin{center}
\epsfig{file=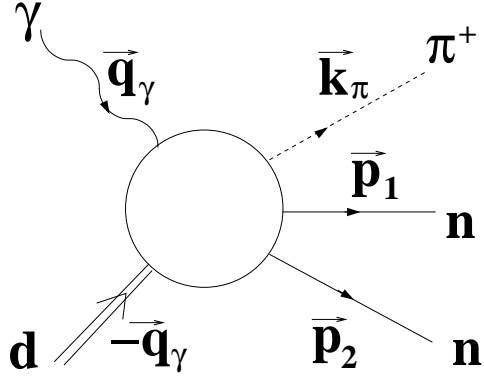, height=5cm, angle=0}
\caption{Definition of the various vectors of the problem.}
\label{vectors}
\end{center}
\end{figure}

\section{Matrix elements and observables}

The differential cross section 
is related to the reaction amplitude via \cite{Berest}

\begin{equation}
d\sigma =(2\pi)^4 \delta^{(4)}(P_f-P_i)\frac12\frac{1}{4q_\gamma
\sqrt{s}}|M_{fi}|^2 
\frac{d^3p_1}{(2\pi)^32E_1}
 \frac{d^3 p_2}{(2\pi)^32E_2} \frac{d^3 k_\pi}{(2\pi)^3 2 \omega_\pi},
\end{equation}
where the indices 1 and 2 label momenta and energies of the final nucleons,
and $k_\pi$ ($\omega_\pi$) denotes the pion three--momentum (energy),
and $M_{fi}$ stands for the sum of all amplitudes given below:
$M_{fi}=M_{a1}+M_{a2}+M_{b1}+\dots$\ \ .
The photon momentum is denoted by $q_\gamma$ and $\sqrt{s}=2M_N+m_{\pi}+Q$ where
$M_N$ ($m_{\pi}$) and $Q$ are the nucleon (pion) mass and the excess energy,
respectively. The choice of variables is illustrated in Fig.~\ref{vectors}.
The factor of 1/2 accounts for the two identical nucleons in the final state.

Let us now present the explicit expressions for the amplitudes corresponding
to the diagrams in Fig.~\ref{diagam}. All calculations are done in the
center--of--mass system of the reaction.  Note that in the calculation of the
leading diagrams (a1) and (a2) in Fig.~\ref {diagam} the D-wave component of
the deuteron wave function was taken into account, whereas in all other
diagrams which are already suppressed by $\chi_m^2$ as compared to the leading
ones, the inclusion of the D-wave is not necessary. This is also true for the diagrams
with $NN$ $P$-wave final state interaction. The loops are evaluated using
nonrelativistic kinematics even for the pion.

The calculation of the diagrams of Fig.\ref{diagam} were done using the
standard Feynman rules with deuteron vertices (for details see, e.g., the
appendix in Ref.\cite{TBK} ).  The explicit expressions  for each individual term are:

\vspace{1cm} \noindent 
{\bf 1. Diagram a1.}

\beqa
&&\hspace{-1cm}M_{a1}= C_{NNLO}\
\chi_1^{\dagger}(\vec \sigma \vec \epsilon_{\gamma})\left[u(\vec q_{2}) (\vec
  \sigma \vec \epsilon_D){-} \frac{w(\vec q_{2})}{\sqrt{2}}
 \left(3\frac{(\vec \sigma \vec q_{2})(\vec q_{2}\vec
      \epsilon_D)}{q_2^2}-(\vec \sigma
    \vec \epsilon_D)\right)\right]\frac{\sigma_2}{\sqrt{2}} \chi_2^* \ - (1
\leftrightarrow 2) \ , \eeqa where
$C_{NNLO}=16\pi\sqrt{M_N}\left(M_N+m_\pi\right) E_{0+}$ ,
$\vec q_{1,2}=\vec p_{1,2}{+} \vec q_\gamma/2$, and $\vec \epsilon_{\gamma}$ and
$\vec\epsilon_D$ are polarization vectors of the photon and the deuteron,
respectively. The notation $- (1\leftrightarrow 2)$ points that one has to
subtract the same term with spin and momentum variables of the two nucleons
interchanged, in order to get properly anti--symmetrized amplitude.

Here $u(\vec p)$ and $w(\vec p)$ are the S-wave and D-wave deuteron
wave functions defined in appendix B.
The expression $\chi_1^{\dagger}\; \hat O\;
\chi_2^* $ corresponds to the spin structure of the final $NN$ pair: 
\beqa
\chi_1^{\dagger}\; (\vec\sigma \vec\epsilon_{\gamma}) (\vec\sigma
\vec\epsilon_D)\frac{\sigma_2}{\sqrt{2}}\; \chi_2^*=(\vec
\epsilon_{\gamma}\vec\epsilon_D)\left( \chi_1^{\dagger}\; \frac{\sigma_2}{\sqrt{2}}
\; \chi_2^* \right) {+} i (\vec\epsilon_{\gamma}\times \vec\epsilon_D)
\left(\chi_1^{\dagger}\; \vec\sigma\frac{\sigma_2}{\sqrt{2}}\; \chi_2^*\right) \eeqa  and
analogously for the D-wave part. Here the first and second terms on the r.h.s.
are the spin-singlet and spin--triplet contributions, respectively
($\chi^{\dagger}\chi=1$).

\vspace{1cm} \noindent 
{\bf 2. Diagram (a2).}

\beqa
\nonumber
M_{a2}\!\!\!&{=}&\!\!\! -C_{NNLO}\
\frac{1}{2M_N}\int \frac{d^3 p}{(2\pi)^3} \left\{\frac{{\cal M}^0_{NN}(\vec p{+}\vec
  k_\pi/2,\vec p_{12},E_{12})}
{p_{12}^2\!-(\vec p{+}\vec k_\pi/2)^2\!+\!i0}u(\vec p{+}\vec q_\gamma/2)\ 
(\vec \epsilon_{\gamma}\vec\epsilon_D)  
\left(\chi_1^{\dagger}\; \frac{\sigma_2}{\sqrt{2}} \; \chi_2^*\right) \right.
\\ 
& & \qquad \qquad \qquad + \left. \frac{{\cal M}^1_{NN}(\vec p{+}\vec
  k_\pi/2,\vec p_{12},E_{12})_{\alpha \beta}}
{p_{12}^2\!-(\vec p{+}\vec k_\pi/2)^2\!+\!i0}u(\vec p{+}\vec q_\gamma/2)\
i (\vec\epsilon_{\gamma}\times  \vec\epsilon_D)_\beta 
\left(\chi_1^{\dagger}\; \sigma_\alpha\frac{\sigma_2}{\sqrt{2}} \; \chi_2^*\right) \right\} \ ,
\eeqa
where $\vec p_{12}=(1/2)(\vec p_1-\vec p_2)$ and the amplitudes
 ${\cal M}^L_{NN}$ are related to the $NN$ partial wave amplitudes with angular
 momentum $L$ as given below in this subsection.
The energy of the $NN$ system is $E_{12}=p_{12}^2/M_N$ .
The formula for $M_{a2}$ as given shows the contribution from the deuteron
$S$--wave only. The inclusion of the $D$--wave in the convolution with $^1S_0$ $NN$ final
state interaction is done like in the expression for diagram (a1). Note that
this diagram is antisymmetrized as well, but in this case the
antisymmetrization reduces just to an additional factor of two, so we have
written the resulting expression explicitly. The same is done for all
diagrams with the half off-shell $NN$ final state interaction, i.e. for the diagrams
(b2), (c2), and (d2), and corresponding diagrams with pion higher partial waves.

The properly anti--symmetrized amplitude ${\cal M}_{NN}$ for the $S-$ and $P$-wave $NN$ interaction with
isospin $1$ in the plane wave basis is
\beqa
\nonumber
\langle \lambda_1'\lambda_2' ,\vec p\, '|M(E)|\lambda_1 \lambda_2 ,\vec p\,\rangle \!\!\!\! &=&\!\!\!\!\!
2\left(\chi_{\lambda_1'}^{\dagger}\; \frac{\sigma_2}{\sqrt{2}} \; \chi_{\lambda_2'}^*\right)
{\cal M}^0_{NN}(\vec  p,\vec p\, ',E)
\left(\chi_{\lambda_1}^T \; \frac{\sigma_2}{\sqrt{2}} \; \chi_{\lambda_2} \right) \\
& & \qquad \quad +
2\left(\chi_{\lambda_1'}^{\dagger}\; \sigma_\alpha\frac{\sigma_2}{\sqrt{2}} \;
  \chi_{\lambda_2'}^*\right) 
{\cal M}^1_{NN}(\vec  p,\vec p\, ',E)_{\alpha \beta}
\left(\chi_{\lambda_2}^T\; \frac{\sigma_2}{\sqrt{2}}\sigma_\beta \; \chi_{\lambda_1} \right) \ ,
\eeqa
where the first (second) term corresponds to the spin--singlet (spin--triplet) $NN$ states.
Here $\lambda'$ and $\lambda$ stand for the spin projections of final and initial nucleons,
respectively, and $\chi_{\lambda}$ for corresponding spinors.
The amplitudes ${\cal M}^L_{NN}$ in this formula are related to the standard $NN$ partial
wave amplitudes $f^{JL}_S$, where $J$ denotes the total angular momentum, 
$L$ the angular momentum 
in the initial and final state (we do not consider couplings between partial
waves with different angular momenta, e.g., the coupling of the $^3P_2$ to the $^3F_2$ partial wave), 
and $S$ the total spin, through 
\beqa
\nonumber
{\cal M}^0_{NN}(\vec  p,\vec p\, ',E) &=& 16\pi M_N f^{00}_0(p',p,E) \\
{\cal M}^1_{NN}(\vec  p,\vec p\, ',E)_{\alpha \beta}&=& 16\pi M_N \sum_J (2J+1)
{\cal   P}^J_{\alpha \beta}
  f^{J1}_1(p',p,E)  \ .
\eeqa

The $f^{JL}_S(p',p,E)$ are related on the mass shell to the partial phase
shifts via

\beqa
\nonumber
f^{JL}_S(p,p,p^2/M_N)=\frac1{2ip}\left(\exp(2i\delta^{JL}_S)-1\right).
\eeqa

The projection operators for the various total angular momenta with $L=L'=1$  and for the $S=1$ read
\beqa
\nonumber
{\cal   P}^0_{\alpha \beta} &=& \hat p'_\alpha
  \hat p_\beta \ ,
 \\ \nonumber
{\cal   P}^1_{\alpha \beta} &=& \frac12 \, \epsilon^{\alpha \lambda \tau}
\epsilon^{\beta \rho \tau}  \hat p'_\lambda \hat p_\rho 
= \frac12\left(\delta_{\alpha \beta}(\hat  p \cdot\hat  p'         )-  \hat
  p'_\beta \hat p_\alpha\right)
\ , \\ \nonumber
{\cal   P}^2_{\alpha \beta} &=& \frac{3}{20}
\left( \delta_{\lambda \alpha}\hat p_\rho ' + \delta_{\rho \alpha}\hat
  p_\lambda '
-\frac23 \delta_{\lambda \rho}\hat p_\alpha '  \right)
\left( \delta_{\lambda \beta} \hat p_\rho  + \delta_{\rho \beta}\hat
  p_\lambda 
-\frac23 \delta_{\lambda \rho}\hat p_\beta   \right)
 \\ \nonumber
 &=&\frac{3}{10}\left(\delta_{\alpha \beta}(\hat p\cdot \hat p ')+  \hat
   p'_\beta \hat p_\alpha
-  \frac23\hat p'_\alpha \hat p_\beta\right) \ ,
\eeqa
where $\hat p=\vec p / |\vec p \, |$ and analogously for $\hat p'$.
These projectors are normalized such that

\beqa
\nonumber
\int d\Omega_{\vec p'}{\cal   P}^J_{\alpha \beta} {\cal   P}^{J'}_{\alpha \beta}&=& 
\frac{4\pi}{2J+1}\ \delta^{JJ'},
\eeqa
and the total neutron-neutron cross-section reads then 
\beqa
\sigma^{nn}_{tot}=\frac12\cdot 4\pi\sum_{JLS}\ (2J+1)|f^{JL}_S(p',p,E)|^2  \ .
\eeqa
Note that the factor of one half occurs in this expression   because the  nucleons  are identical.

\vspace{1cm} \noindent 
{\bf 3. Diagram (b1). }

The real part of the diagram (b1) renormalizes the bare vertex in the
leading diagram (a1) resulting in  the experimentally observed value
$E_0^+$ for $\gamma p\to \pi^+ n$ process. Thus, the real part of this
diagram is already included in the expression for diagram (a1). 
The imaginary part of (b1) is 
\beqa
\hspace{-1.4cm}M_{b1}= i C_{LO} \frac{m_{\pi}}{8\pi(1+m_{\pi}/M_N) f^2_{\pi}}
\ u(\vec p_{2}\!{+}\!\vec q_\gamma/2)\  k_{\pi N_1}\chi_1^{\dagger} (\vec \sigma \vec \epsilon_{\gamma})
(\vec \sigma \vec \epsilon_D)\frac{\sigma_2}{\sqrt{2}} \chi_2^* - (1 \leftrightarrow  2),
\eeqa
where $k_{\pi N_i}{=}|M_N\vec k_\pi {-} m_\pi \vec p_i|/(M_N{+}m_\pi)$ is the
magnitude of  the relative momentum of the final pion and the final nucleon
with momentum $p_i$ and $C_{LO} = 4\sqrt{M_N}g_{\pi N}e/(\sqrt{2})$.

\vspace{1cm} \noindent 
{\bf 4. Diagram (c1).}


\beqa
\hspace{-1cm}M_{c1}=- C_{LO}\ \frac{m_{\pi}}{2(1+m_{\pi}/M_N)f^2_{\pi}}
\; I_{c_1} \;
 \chi_1^{\dagger} (\vec \sigma \vec \epsilon_{\gamma})
(\vec \sigma \vec \epsilon_D)\frac{\sigma_2}{\sqrt{2}} \chi_2^* \ - (1 \leftrightarrow  2),
\eeqa
where the integral $I_{c_1}$ is
\beqa 
I_{c_1}=\int \frac{d^3 p }{(2\pi)^3} \
\frac{u(\vec p\!+\!\vec q_\gamma/2)}{ k_{\pi N_1}^2- (\vec p + \frac{M_N}{M_N+m_{\pi}}\vec p_1)^2+i0} \ .
\eeqa

\newpage

\vspace{1cm} \noindent 
{\bf 5. Diagram (b2)}

\beqa
M_{b_2}= C_{LO} \
\frac{1}{8M_Nf_{\pi}^2} I_{b_2}
(\vec \epsilon_{\gamma}\vec\epsilon_D)  
\left(\chi_1^{\dagger}\; \frac{\sigma_2}{\sqrt{2}} \; \chi_2^*\right) \ , 
\eeqa
with 
\beqa
I_{b_2}=\int \frac{d^3 p d^3 l}{(2\pi)^6} \frac{{\cal M}^0_{NN}(\vec p{+}\vec
  k_\pi/2,\vec p_{12},E_{12})}
{p^2_{12}{-}(\vec p{+}\vec k_\pi/2)^2{+}i0}\;
\frac{u(\vec p{+}\vec q_\gamma/2)}{Q{-}\frac{p^2}{2M_N}{-}\frac{l^2}{2M_N}
{-}\frac{(\vec l{+} \vec p)^2}{2m_{\pi}}{+}i0}
\label{ib2}
\eeqa
The integral with the $\pi N$ loop, i.e. over $d^3 l$, is divergent and
has to be renormalized (cf. discussion of diagram (b1)). After renormalization it takes the form
\beqa
I_{b_2}=\frac{-i m_{\pi}}{2\pi(1{+}m_{\pi}/M_N)}  \int
\frac{d^3 p}{(2\pi)^3} 
 \frac{{\cal M}^0_{NN}(\vec p{+}\vec
  k_\pi/2,\vec p_{12},E_{12})}
{p^2_{12}{-}(\vec p{+}\vec k_\pi/2)^2{+}i0}\;
u(\vec p{+}\vec q_\gamma/2) \; K(Q,p) \ ,
\eeqa
where $m_{\pi N}=m_{\pi}M_N/(M_N{+}m_{\pi})$,
$\mu_N=M_N(M_N{+}m_{\pi})/(2M_N{+}m_{\pi})$, 
and  $$K(Q,p)=\sqrt{2m_{\pi N}(Q{-} p^2/2\mu_N)} \ .$$
Note that for negative arguments the square root needs to be replaced
by $i\sqrt{2m_{\pi N}(p^2/2\mu_N{-}Q)}$. Thus, we get a purely real,
non--vanishing contribution from
diagram (b2) even at production threshold.

\vspace{1cm} \noindent 
{\bf 6.  Diagram (c2)}

\beqa
M_{c_2}= C_{LO}\
\frac{1}{8M_Nf_{\pi}^2} I_{c_2}
(\vec \epsilon_{\gamma}\vec\epsilon_D)  
\left(\chi_1^{\dagger}\; \frac{\sigma_2}{\sqrt{2}} \; \chi_2^*\right) \ , 
\eeqa
with 
\beqa
I_{c_2}=\int \frac{d^3 p d^3 l}{(2\pi)^6} \frac{{\cal M}^0_{NN}(\vec p{+}\vec
  k_\pi/2,\vec p_{12},E_{12})}{p^2_{12}{-}(\vec p{+}\vec k_\pi/2)^2{+}i0}\;
\frac{u(\vec l{+}\vec  q_\gamma/2)}{Q{-}\frac{p^2}{2M_N}{-}\frac{l^2}{2M_N}{-}\frac{(\vec p{+} \vec l)^2}{2m_{\pi}}{+}i0}.
\label{ic2}
\eeqa
The integral $I_{c_2}$ develops an angle dependent three--body singularity as soon as we move
away from the production threshold. Fortunately, for non--relativistic pions,
it transformed into an angle independent one. Indeed, rewriting the
denominator in $I_{c_2}$ that creates the three--body singularity, as follows
\beqa
\nonumber
Q{-}\frac{p^2}{2M_N}{-}\frac{l^2}{2M_N}{-}\frac{(\vec p{+} \vec
  l)^2}{2m_{\pi}}=Q{-}\frac{p^2}{2\mu_N}{-}\frac{(\vec l{+}\frac{m_{\pi N}}{m_\pi}\ \vec
  p)^2}{2m_{\pi N}},
\eeqa
and shifting the integration variable $\vec l$ by $\frac{m_{\pi N}}{m_\pi}\ \vec
  p$, we get the integral over $d^3 l$ in Eq.~(\ref{ic2}) as

\begin{equation}
\nonumber
\int\frac{d^3l}{(2\pi)^3}\frac{u(\vec l+\vec  q_\gamma/2)}
{Q-\frac{l^2}{2M_N}-\frac{p^2}{2M_N}-\frac{(\vec l+\vec
    p)^2}{2m_\pi}+i0}=2m_{\pi N} \int\frac{d^3l}{(2\pi)^3}\frac{u(\vec l+\vec  q_\gamma/2-\frac{m_{\pi N}}{m_\pi}\ \vec
  p)}{K^2(Q,p)-l^2+i0}\ ,
\end{equation}
from which one can immediately see that the moving three--body singularity
turned to the frozen one -- the pole position does not depend on $\vec l$ angles.
This is possible, because in our reaction the initial two
nucleons are always off the mass shell -- they form a bound
state. Therefore, there is no additional propagator which can cause a
singularity.

Furthermore, for non--relativistic pions and specific analytical parameterizations of
the deuteron wave functions, the three body singuilarity can be integrated
analytically. In particular, using the parameterization for the CD-Bonn 
deuteron wave function \cite{CDBonn} (cf. appendix B) we may write 
\begin{equation}
\int\frac{d^3l}{(2\pi)^3}\frac{u(\vec l+\vec  q_\gamma/2)}
{Q-\frac{l^2}{2M_N}-\frac{p^2}{2M_N}-\frac{(\vec l+\vec
    p)^2}{2m_\pi}+i0}=-\frac{2 m_\pi}{1+m_\pi/M_N}\sum\limits_j C_j \frac{1}{8\pi i r}
\log\left(\frac{\beta_j-iK(Q,p)+ir}{\beta_j-iK(Q,p)-ir}\right) \ ,
\label{3bs}
\end{equation}
where $\vec r=\frac{1}{1+m_\pi/M_N}\vec p-\vec q_\gamma/2$. The integration in
Eq.~(\ref{3bs}) has been performed similarly to the case of diagram $(a2)$: we
applied an inverse Fourier transform separately to the deuteron wave function
and to the Green's function. Then the integration over $d^3l$ gives a
delta--function, which kills one of the remaining three--dimensional
integrations and the integral reduces to a sum of integrals of the type
considered, e.g., in Ref.\cite{bkt}.  The three--body singularities of
the diagrams (d1) and (d2) can be handled analogously.  Also for the $NN$
 final state interaction the integration over one of the loops can
be performed analytically after applying the procedure outlined in Refs.
\cite{Johann,Johann2} to the $NN$ amplitudes, for they can then be represented in the same
analytical form as the deuteron wave functions, cf.  appendix B for details.
Moreover, it turns out that in diagram (d2) one can perform analytically two
of the three loop integrations, thus reducing the nine--dimensional integral
to the three--dimensional one.

\vspace{1cm} \noindent 
{\bf 7.  Diagram (d1)}

\beqa
M_{d_1}= C_{LO}\
\frac{1}{8M_Nf_{\pi}^2} I_{d_1}
(\vec \epsilon_{\gamma}\vec\epsilon_D)  
\left(\chi_1^{\dagger}\; \frac{\sigma_2}{\sqrt{2}} \; \chi_2^*\right)
 {-} (1 \leftrightarrow  2),
\eeqa
with 
\beqa
I_{d_1}=\int \frac{d^3 p d^3 s}{(2\pi)^6} \frac{{\cal M}^0_{NN}(\vec
  p{+}\vec  s/2,\vec p_{2}{+}\vec s/2,E_{NN})}{M_NE_{NN}{-}(\vec p_2+\vec  s/2)^2{+}i0}
\frac{u(\vec p{+}\vec
  q/2)}{Q{-}\frac{p^2}{2M_N}{-}\frac{s^2}{2m_{\pi}}{-}\frac{(\vec p{+} \vec
    s)^2}{2M_N}{+}i0} \, ,
\eeqa
where $E_{NN}=Q{-}s^2/2\mu_{\pi}$ with $\mu_{\pi}=2M_Nm_{\pi}/(2M_N{+}m_{\pi})$.

\vspace{1cm} \noindent 
{\bf 8.  Diagram (d2)}

\beqa
M_{d_2}= C_{LO}\
\frac{1}{16M_Nf_{\pi}^2} I_{d_2}
(\vec \epsilon_{\gamma}\vec\epsilon_D)  
\left(\chi_1^{\dagger}\; \frac{\sigma_2}{\sqrt{2}} \; \chi_2^*\right) \ , 
\eeqa
with 
\beqa
\nonumber
I_{d_2}&=&-\frac{1}{M_N}\int \frac{d^3 l\; d^3 s\; d^3p}{(2\pi)^9} 
\frac{{\cal M}^0_{NN}(\vec p{+}\vec k_\pi/2,\vec p_{12},E_{12})}
{p_{12}^2{-}(\vec p{+} \vec k_\pi/2)^2{+}i0}\\
&\times&
\frac{{\cal M}^0_{NN}(\vec l{+}\vec s/2,\vec p{+}\vec s/2,E_{NN})}
{M_NE_{NN}{-}(\vec p{+} \vec s/2)^2{+}i0}
\frac{u(\vec l{+}\vec
  q/2)}{Q{-}\frac{l^2}{2M_N}{-}\frac{s^2}{2m_{\pi}}{-}\frac{(\vec l{+} \vec
    s)^2}{2M_N}{+}i0} \, .
\eeqa


\vspace{1cm} \noindent 
{\bf 9.  Corrections to  diagrams (a1) and (a2) from higher pion partial waves.}

\vspace*{0.5cm} 

We start from the explicit expression for the pion $p$--wave
contribution (below labeled as $(\pi{-}p)$) stemming from diagram (b) of Fig.
\ref{pwaves}. To implement this, the term $(\vec \sigma \vec
\epsilon_{\gamma})$ in the expressions for diagram (a1) and (a2) of Fig.
\ref{diagam} (see above) needs to be replaced by $ (\vec \sigma \hat q) (\vec
\epsilon_\gamma \vec k_\pi)/m_\pi \ , $ where $\hat q=\vec q_\gamma / |\vec q_\gamma \, |$.
Thus we get 
\beqa 
\nonumber &&\hspace{-1cm}M_{a1}^{(\pi{-}p)}{=} C_{LO}\ 
\frac1{m_\pi}(\vec \epsilon_\gamma \vec k_\pi) \chi_1^{\dagger} (\vec \sigma
\hat q) (\vec \sigma \vec \epsilon_D)\frac{\sigma_2}{\sqrt{2}} \chi_2^* \ 
u(\vec q_{2}) \ - (1 \leftrightarrow 2) \ \eeqa for the diagram without final
state interaction and \beqa \nonumber M_{a2}^{(\pi{-}p)}{=} -C_{LO}\ 
\frac1{m_\pi}(\vec \epsilon_\gamma \vec k_\pi) \frac{1}{2M_N}\int \frac{d^3
  p}{(2\pi)^3} \frac{{\cal M}^0_{NN}(\vec p{+}\vec k_\pi/2,\vec p_{12},E_{12})}
{p_{12}^2\!-(\vec p{+}\vec k_\pi/2)^2\!+\!i0}u(\vec p{+}\vec q_\gamma/2)\ (\hat q
\vec\epsilon_D) \left(\chi_1^{\dagger}\; \frac{\sigma_2}{\sqrt{2}} \;
  \chi_2^*\right) 
\eeqa 
for the diagram with $NN$ final state interaction.

Note, that when the pion is in a $p$--wave, only those terms where the $NN$
final state is in an $S$--wave are to be considered.  The
simultaneous appearance of two $p$--waves in the final state is strongly
suppressed by the centrifugal barrier.
Our numerical calculations confirm this  statement.


Using the vertex 
$V_{\pi\gamma NN}^{(c)}$ given by
Eq.~(\ref{magmom}) as  input for the diagrams (a1) and (a2) of
Fig.~\ref{diagam}  one can get the corresponding contribution from the
s- and u-channel nucleon pole diagrams (cf. diagrams (c) in Fig.~\ref{pwaves}) to  our reaction as follows
\beq
M^{nuc-su}=-\frac{C_{LO}}{2m_{\pi}M_N}
\left \{(A^{nuc-su}+F^{nuc-su})
\left(\chi_1^{\dagger}\; \frac{\sigma_2}{\sqrt{2}} \; \chi_2^*\right)
+\vec B^{nuc-su}\left(\chi_1^{\dagger}\; \vec\sigma \frac{
  \sigma_2}{\sqrt{2}} \; \chi_2^*\right) \right\},
\eeq
where
\beqa
A^{nuc-su}= u(\vec q_2)
\left \{
2(\vec  \epsilon_{\gamma}\vec p_2) (\vec k_{\pi} \vec\epsilon_D) {+}
(\mu_p+\mu_n)\left((\vec \epsilon_\gamma \vec k_\pi)(\vec q_\gamma \vec \epsilon_D)  {-}
(\vec q_\gamma \vec k_\pi)(\vec \epsilon_\gamma \vec\epsilon_D)\right)\right \}
{+} (1 \leftrightarrow   2),
\eeqa

\beqa
\nonumber
\hspace*{-5.5cm}\vec B^{nuc-su}&=& u(\vec q_2)\left \{
2(\vec  \epsilon_{\gamma}\vec p_2) (\vec k_{\pi}\times \vec\epsilon_D) +
(\mu_p-\mu_n)(\vec k_\pi (\vec \epsilon_\gamma\times \vec q_\gamma))
\vec \epsilon_D \right.\\
\hspace*{-1.4cm} 
&+& \left. (\mu_p +\mu_n)\left(
(\vec \epsilon_\gamma \vec k_\pi)(\vec q_\gamma\times\vec \epsilon_D) -  
(\vec q_\gamma \vec k_\pi)(\vec \epsilon_\gamma\times \vec \epsilon_D)
\right)
\right \}
- (1 \leftrightarrow  2),
\eeqa

and $F^{nuc-su}$ stands for the contributions from the diagrams with $NN$
final state interaction:

\beqa
\nonumber
\hspace*{-2.0cm} F^{nuc-su}&=&(\mu_p+\mu_n)\left((\vec \epsilon_\gamma \vec k_\pi)(\vec q_\gamma \vec \epsilon_D)  {-}
(\vec q_\gamma \vec k_\pi)(\vec \epsilon_\gamma \vec\epsilon_D)\right)F_1-(\vec k_\pi\vec \epsilon_\gamma)(\vec k_\pi\vec \epsilon_D)F_2,
\eeqa

where
\beqa
\nonumber
F_1&=& -\frac{1}{2M_N}\int \frac{d^3 p}{(2\pi)^3} \frac{{\cal M}^0_{NN}(\vec p{+}\vec
k_\pi/2,\vec p_{12},E_{12})}{p_{12}^2\!-(\vec p{+}\vec
k_\pi/2)^2\!+\!i0}u(\vec p{+}\vec q_\gamma/2),\\
\nonumber
F_2&=& -\frac{1}{2M_N}\int \frac{d^3 p}{(2\pi)^3} \frac{{\cal M}^0_{NN}(\vec p{+}\vec
k_\pi/2,\vec p_{12},E_{12})}{p_{12}^2\!-(\vec p{+}\vec
k_\pi/2)^2\!+\!i0}u(\vec p{+}\vec q_\gamma/2)\left(1-\frac{(\vec p+\vec k_\pi/2)(\vec
k_\pi/2-\vec q_\gamma/2)}{(\vec
k_\pi/2-\vec q_\gamma/2)^2}\right).
\eeqa

\section{The $NN$ wave functions} 

For the deuteron wave function, and for the $nn$
scattering amplitudes that appear in the final-state interaction,
we take those of the (charge dependend) CD-Bonn $NN$ potential \cite{CDBonn}.
In particular, we
utilize the analytic parameterization of the deuteron 
wave function provided in Ref. \cite{CDBonn} which is given
by
\beq
u(p)=\sqrt{4\pi}\sum_j C_j/(p^2+m_j^2); \ w(p)=\sqrt{4\pi}\sum_j D_j/(p^2+m_j^2), 
\eeq
with parameters listed in Table 20 of that reference. The wave function is normalized according to 
\beqa \int \frac{d^3 p}{(2\pi)^3}\;
(u^2(p){+}w^2(p))= 1 \, .  \eeqa
With this parameterization some of the diagrams can be evaluated
analytically. 
In order to facilitate also an analytic evaluation of the diagrams involving 
the $nn$ scattering amplitude the CD-Bonn potential in the relevant partial 
waves ($^1S_0$, $^3P_0$, $^3P_1$, $^3P_2$) is cast into a separable 
representation by means of the so-called EST method \cite{Johann2}.  
The resulting rank-1 separable interactions exactly reproduce the 
on- and off-shell properties of the CD-Bonn potential at the
chosen approximation energies ($E_{Lab}= 0$ MeV for $^1S_0$ and 
$E_{Lab}= 30$ MeV for the $P$ waves) \cite{Johann2} and they provide also an
excellent approximation in a broad neighborhood of these energies. 
The form factors $g(p)$ of these separable representations, that consist
of the scattering solutions of the CD-Bonn potential at the
specified approximation energies \cite{Johann2}, 
are parameterized in analytical form, 
\beq
g(p)=\sum_i c_i/(p^2+\beta_i^2), 
\eeq
for $^1S_0$ and 
\beq
g(p)=\sum_i c_i p/(p^2+\beta_i^2)^2,
\eeq
for the $P$ waves and the scattering amplitude is then given by
\beq
f(p,p';k) = \frac{2\pi^2 M_N g(p)g(p')}{\pm 1 - M_N \int d^3 q \frac{g^2(q)}{q^2-k^2-i0}} .
\eeq
Here the positive sign pertains to the $^1S_0$, $^3P_0$, and $^3P_2$ partial
waves, and the negative sign to the $^3P_1$ partial wave.
The parameters $c_i$ and $\beta_i$ for each partial wave
are listed in Table \ref{separable}.

\begin{table}[ht]
\begin{center}
\begin{tabular}{rcccc}
\hline
\hline
&\multicolumn{2}{c}{$^1S_0$}&\multicolumn{2}{c}{$^3P_0$}\\
\hline
& $c_i$ [MeV] & $\beta_i$ [MeV] & $c_i$ [MeV$^2$]& $\beta_i$ [MeV] \\
\hline
1 &-1.6788489  &105.64868 &5.6091364  &87.697924 \\
2 & 38.388276  &208.40749 &-47.29225  &196.68429  \\
3 &-204.19687  &311.16630 &-52680.52  &305.67065 \\
4 &-265.30647  &413.92511 &25403.007  &414.65702 \\
5 & 604.93218  &516.68392 &140524.04  &523.64338  \\
\hline
\hline
&\multicolumn{2}{c}{$^3P_1$}&\multicolumn{2}{c}{$^3P_2$}\\
\hline
& $c_i$ [MeV$^2$]& $\beta_i $[MeV] & $c_i$ [MeV$^2$] & $\beta_i$ [MeV]\\
\hline
1 &-26.122635  &151.92170  &-169.35853  &139.80616  \\
2 & 107984.65  &329.53539  & 12651.165  &243.67922  \\
3 &-1107050.3  &507.14909  &-149453.00  &347.55228 \\
4 & 3691951.8  &684.76279  & 405215.24  &451.42534 \\
5 &-3372798.2  &862.37648  &-389168.49  &555.29840  \\
\hline
\end{tabular}
\caption{Parameters of the form factors for the separable representation 
of the CD-Bonn potential. 
}
\label{separable}
\end{center}
\end{table}

\end{document}